\documentclass[11pt]{article}
\usepackage{amsmath}
\usepackage{amssymb}
\usepackage[breaklinks,colorlinks=true,linkcolor=blue,citecolor=red,backref=page]{hyperref}
\usepackage{graphicx}
\graphicspath{ {./images/} }
\usepackage{placeins}
\usepackage{float}
\usepackage{subcaption}

\DeclareMathAlphabet{\itbf}{OML}{cmm}{b}{it}

\newcommand{\ba}{{\itbf a}}
\newcommand{\bb}{{\itbf b}}
\newcommand{\bc}{{\itbf c}}

\newcommand{\bw}{{\itbf w}}
\newcommand{\bZ}{{\itbf Z}}
\newcommand{\bX}{{\itbf X}}

\newcommand{\balpha}{{\boldsymbol \alpha}}
\newcommand{\bzeta}{{\boldsymbol \zeta}}

\newcommand{\eps}{\varepsilon}

\newcommand{\PP}{{\mathbb P}}
\newcommand{\EE}{{\mathbb E}}

\def\qed{\hbox{${\vcenter{\vbox{  
   \hrule height 0.4pt\hbox{\vrule width 0.4pt height 6pt
   \kern5pt\vrule width 0.4pt}\hrule height 0.4pt}}}$}}

\begin{document}

\title{The Climate Extended Risk Model (CERM)}

\date{\today}
\author{Josselin Garnier\footnote{
CMAP,~Ecole~polytechnique, Institut Polytechnique de Paris, 91128 Palaiseau Cedex, France,
%~https://www.josselin-garnier.org; 
and 
%\hspace*{0.4in}
Lusenn, 73 rue L\'eon Bourgeois, 91120 Palaiseau, France
%,~http://www.lusenn.com/
},
Jean-Baptiste Gaudemet\footnote{Green RWA, 14 rue Sainte Adelaide 78000 Versailles, France and et Amalthea FS, 7 rue Benjamin Franklin 75116 Paris, France
},
and
Anne Gruz\footnote{
Iggaak, 11 rue René Blum, 75017 Paris, France
%~https://www.iggaak.com; 
}
}
\maketitle

\begin{abstract}
This paper addresses estimates of climate risk embedded within a bank credit portfolio. The proposed Climate Extended Risk Model (CERM) adapts well known credit risk models and makes it possible to calculate incremental credit losses on a loan portfolio that are rooted into physical and transition risks. The paper provides detailed description of the model hypotheses and steps.
\end{abstract}

% 	91G40    Credit risk
%     91G10    Portfolio theory
% 60H10   	Stochastic ordinary differential equations (aspects of stochastic analysis) 

\section{Introduction}
During the 21st century, man-made green house gases (GHGs) emissions will raise global temperatures, resulting in severe and unpredictable physical damage across the globe. Another uncertainty associated with climate, known as the energy transition risk, comes from the unpredictable pace of political and legal actions to limit this impact.
Both risks will impact the risk assessment of bank loan portfolios. 
Credit risk modeling and assessment have been studied for a long time. An extensive review can be found in \cite{roncalli}.
It is also subjected to regulations \cite{basel,basel3}.
Those models and regulations take into account the economic risk but neglect -so far- the physical and transition risks.
The goal of this paper is to assess the loss of a credit portfolio under stressed conditions taking into account 
economic, physical and transition risks. The idea is to develop a model that is a direct extension of the models used in the current regulations. 
Within a bank the model can be used internally to manage capital buffers, as well as in a regulatory context for Pillar 1 and Pillar 2 requirements of the Basel regulatory framework. Within an investing company it can be used to better assess the risk/return relationship.
Even-though we refer to loans hereafter, the model applies to debt instruments in general, in particular to loans and to bonds.

A loan portfolio comprises loans from a large number of borrowers, made up of different groups representing geographic regions, and/or economic sectors, and/or climate risk mitigation and adaptation strategies, and/or collateral types and which have different ratings\footnote{Credit ratings are external and depend on credit rating agencies. The most famous are Standard \& Poor's, Moody's and Fitch Ratings.} at the initial time.
The individual losses of the borrowers are random and may depend  on systematic risk factors and on idiosyncratic risk factors.
The systematic risk factors model economic, transition, and physical risks. They may be correlated and they influence all borrowers.
The idiosyncratic risk factors are specific to each borrower and they are independent from each other (and independent from the systematic risk factors).
The loss of a portfolio is the sum of the random losses of the borrowers. 
It is, therefore, random. We are typically interested in the expected loss and in the loss quantiles.

The expected loss of the portfolio is simply the sum of the expected individual losses. There is no need to describe the dependence structure between the individual losses to get the expected loss of the portfolio.
As a result, the expected loss can be expressed in terms of probability of default, exposure at default, and average loss given default.
The probability of default can be obtained from unconditional migration matrices. Those matrices express the probabilities for one borrower to move from one rating to another one (until the ultimate rating which corresponds to default) during a unit time interval.
They are unconditional in the sense that they are averaged over idiosyncratic and systematic risk factors.
The loss given default can be based on a deterministic or random recovery model that also depends on idiosyncratic and systematic risk factors. 
The correlation between default occurrence and recovery rate is only through the systematic risk factors. 
The exposure at default (given default at a certain time) is deterministic and independent from the migration and recovery processes. It can model various banking portfolio dynamics (flat, amortizing...).

The unexpected loss is defined as a quantile of the loss of the portfolio (value at risk). In contrast with the expected loss that is the sum of the expected individual losses, the quantile of the loss of the portfolio cannot be expressed in terms of the quantiles of the individual losses, because the quantile of a sum is not the sum of the quantiles. A model is, therefore, necessary to describe the dependence structure between the random losses of the borrowers.

The Asymptotic Single Risk Factor (ASRF) model is a default-mode (Merton-type) model proposed by Vasicek in 1991 \cite{vasicek91}. It has played a central role for its regulatory applications in the Basel Capital Adequacy Framework (BCAF) \cite{basel}. 
The Basel IRB (Internal ratings-based) ASRF model 
calculates the loss conditional to a single systematic economic risk factor.
It is based on the following assumptions:\\
- a unique systematic risk factor (single-factor model): economic risk, so that the losses of the borrowers are correlated only through one systematic factor,\\
- an infinitely granular portfolio characterized by a large number of small size loans, so that  the idiosyncratic risks are diversified, but not the systematic risk,\\
- a dependence structure by a Gaussian copula.\\ 
The ASRF model can be used to express unconditional and conditional migration matrices and it can give closed-form expressions for the expected and unexpected losses \cite{belkin98,roncalli}.

The Climate-extended risk model (CERM) introduced in this paper is a multi-factor Merton-type model \cite{pykhtin04}.
It is based on the following assumptions:\\
- several systematic risk factors (multi-factor model): economic, physical, transition risks,\\
- an infinitely granular portfolio,\\
- a dependence structure described by a Gaussian copula.\\
The CERM makes it possible to build efficient and rapid Monte-Carlo estimations of the expected and unexpected losses.
Its definition is constrained by the fact that we wish to design a model that:\\
- is an extension of the ASRF model and should not give rise to any discontinuity when switching from the version used in the BCAF or derived from the BCAF,\\
 - can be calibrated given a comprehensive rating system that  provides the relative exposures of borrowers to each systematic risk,\\
- makes it possible to estimate portfolio loss distributions conditional to a physical and transition scenario (temperature rise and GHGs pathway). \\ 
The main challenge facing the CERM compared to classical  multi-factor Merton-type models is to address non-stationarities due to the evolutions of the physical and transition risk intensities.
In the CERM the physical and transition scenario gives the evolution of the physical and transition systematic risk intensities.
They can be based on scenarios proposed by the Network for Greening the Financial System (NGFS, a group of 100 central banks and supervisors) or by the International Energy Agency (IEA), both derived from the work done by the Intergovernmental Panel on Climate Change
(IPCC\footnote{https://www.ipcc.ch/})
(NGFS\footnote{https://www.ngfs.net/})
(IEA\footnote{https://www.iea.org/}).
We would like to consider different scenarios such as
``Stated Policies Scenario",
``Announced Pledges Scenario",
``Sustainable Development Scenario", 
``Net Zero Scenario" in the IEA report and to show that they may have very different impacts on the expected and unexpected losses, so as to motivate the relevance of political decision.

The paper is organized as follows. 
In Section \ref{sec:portloss} we present the portfolio loss.
In Section \ref{sec:exploss} we give the expected loss of a portfolio.
In Section \ref{sec:model} we present the model for the default occurrence and the recovery rate that is used to determine the dependence structure between the individual losses.
In Section \ref{sec:condloss} we give the conditional loss of a portfolio under stressed conditions for a given systematic risk trajectory.
We specify the model for the systematic risk factors in Section \ref{sec:riskfactors}.
 We describe the calibration of the risk factors using dynamic macro and micro correlations coming from climate models and climate analysts in Section \ref{sec:loading}.
 We propose a sensitivity and risk allocation analysis in Section \ref{sec:ana}.
We give some additional perspectives in Section  \ref{sec:rev} 
and we present an application to a pilot portfolio in Section \ref{sec:app}.

\section{Portfolio loss}
\label{sec:portloss}
The loss of a loan portfolio of $N$ loans can be written as:
$$
L = \sum_{t=1}^T \sum_{q=1}^N l^{(q)}_{t}  \chi^{(q)}_{t}  .
$$
The time is discretized as integers $t=0,\ldots,T$, where $t=0$ is present and $t=T$ is the time horizon of the stress test analysis.
$\chi^{(q)}_{t} $ is an indicator function that is equal to one if the $q$-th borrower defaults at time $t$ and $0$ otherwise.
The loss is then denoted by $l^{(q)}_{t}$
and has the form
\begin{align}
\label{eq:loss}
l^{(q)}_{t} &= {\rm EAD}^{(q)}_t \big[ 1 -{\rm RR}^{(q)}_{t} \big],
\end{align}
where 
\begin{itemize}
\item
${\rm EAD}^{(q)}_t $ is the exposure (the total balance owed by the borrower at time of default) of the $q$-th borrower given default at time $t$,
\item
${\rm RR}^{(q)}_{t}$ is the recovery rate (the proportion of the exposure that is recovered by way of liquidation of collateral and other resolution or post-default collection actions) of the $q$-th borrower given default at time $t$. The loss given default (the proportion of the exposure that is lost if the borrower defaults) is $1-{\rm RR}^{(q)}_{t}$.
\end{itemize}

The exposure at default is deterministic and independent from the default and recovery processes.
It is determined by the principal 
%(nominal amount)
 and the amortization profile of the loan.
For instance, for an amortizing loan with principal $K^{(q)}$, maturity $T^{(q)}$, interest rate $r^{(q)}$ and equal payments, we have 
\begin{equation}
{\rm EAD}^{(q)}_t = K^{(q)}
\frac{ (1+r^{(q)})^{T^{(q)}}-(1+r^{(q)})^t}{(1+r^{(q)})^{T^{(q)}}-1} {\bf 1}_{t \leq T^{(q)}}
.
\end{equation}
Note that $T^{(q)}$ can be larger or smaller than $T$,  which means there is no constraint on the distribution of the loan maturities.

Each borrower belongs to a group and has a rating.
We assume that:
\begin{itemize}
\item
There are $G$ groups. A group can represent a geographic region, and/or an economic sector, and/or a climate risk mitigation and adaptation strategy\footnote{
Climate change mitigation consists of actions to lessen the magnitude or the rate of global warming and its related effects. This generally involves reductions in emissions of greenhouse gases.
Climate change adaptation consists of incremental adaptation actions where the central aim is to maintain the essence and integrity of a system or of transformational adaptation actions that change the fundamental attributes of a system in response to climate change and its impacts.} and/or a collateral type.
\item
There are $K$ rating levels $\{1,\ldots,K\}$. The rating $K$ corresponds to default.
\end{itemize}

\section{Expected loss}
\label{sec:exploss}
The expected loss of the portfolio is the sum of the expected individual losses. By grouping these terms into groups, it is given by
\begin{align}
\label{def:lossexp}
L^{\rm e} =&  \sum_{t=1}^T L^{\rm e}_t , \\
\label{def:lossexp1}
L^{\rm e}_1=& \sum_{{g}=1}^{G} \sum_{i=1}^{K-1} 
({\bf M}_{{g},1})_{iK} {\rm LGD}_{{g},i,1}^{\rm e} {\rm EAD}_{{g},i,1} , \\
L^{\rm e}_t=&\sum_{{g}=1}^{G} \sum_{i,j=1}^{K-1} 
\big( {\bf M}_{{g},1}\cdots{\bf M}_{{g},t-1} \big)_{ij} ({\bf M}_{{g},t})_{jK}
 {\rm LGD}_{{g},j,t}^{\rm e} {\rm EAD}_{{g},i,t}
 ,\quad t\geq 2,
\label{def:lossexpt}
\end{align}
where:
\begin{itemize}
\item
$L^{\rm e}_t$ is the expected loss due to the defaults that occur at time $t$.
The first term $L^{\rm e}_1$ in (\ref{def:lossexp}) is the loss due to the borrowers who default at time $1$. It is decomposed in (\ref{def:lossexp1}) over all possible groups $g$ and initial rating $i$.
The $t$-th term $L^{\rm e}_t$ in (\ref{def:lossexp}) is the loss due to the borrowers who default at time $t$. It is decomposed in (\ref{def:lossexpt}) over all possible groups $g$, initial rating $i$, and rating $j <K$ that the borrowers may have at time $t-1$.
\item
$ {\rm EAD}_{{g},i,t}$ is the total exposure at default, given default at time $t$ due to the borrowers in group ${g}$ and with initial rating $i$:
\begin{equation}
\label{def:wgi}
 {\rm EAD}_{{g},i,t}
 = \sum_{q=1}^N {\rm EAD}^{(q)}_t {\bf 1}_{\mbox{\small $q$-th borrower is in group ${g}$ and has initial rating $i$}} .
\end{equation}
${\rm EAD}_{{g},i,t}$ can be seen as the maximal loss at time $t$ from the borrowers in group ${g}$ and with initial rating $i$ (in the worst case scenario when they all default at time $t$ with zero recovery rate). 
\item
${\bf M}_{{g},t}$ is the unconditional migration matrix (of size $K\times K$) at time $t$; $({\bf M}_{{g},t})_{ij}$ is the probability for a borrower in group ${g}$ and with rating $i$ at time $t-1$ to migrate to rating $j$ at time $t$ (see Subsection \ref{subsec:umm}). 
In particular the $i$-th entry of the last column $({\bf M}_{{g},t})_{iK}$ gives the probability of default at time $t$ for a borrower in group ${g}$ and with rating $i \in \{ 1,\ldots,K-1 \}$ at time $t-1$.\\
$\big( {\bf M}_{{g},1}\cdots{\bf M}_{{g},t-1} \big)_{ij} ({\bf M}_{{g},t})_{jK}$ can be interpreted as the probability that a borrower in group $g$ and with rating $i$ at time zero has rating $j$ at time $t-1$ and defaults at time $t$.
If we sum over $j=1,\ldots,K-1$, we obtain the probability that a borrower in group $g$ and with rating $i$ at time zero defaults at time $t$ (and not before).
\item
${\rm LGD}_{{g},j,t}^{\rm e}$ is the average Loss Given Default for a borrower in group ${g}$ and with rating $j$ at time $t-1$ who defaults at time $t$ (its rating jumps from $j$ to $K$) (see Subsection \ref{subsec:lgd}).
\end{itemize}

The framework proposed in this paper can be used when new loans are added to the portfolio at different times.
Indeed, if new loans are added at time $t_0>0$,
then they can be incorporated into the model by creating new groups $g'$ which are such that the migration matrices ${\bf M}_{{g}',t}$ are equal to the identity matrix ${\bf I}$ for $t \leq t_0$. 
This would make it possible to address various dynamic balance sheet strategies (where the composition or risk profile of the portfolio is allowed to vary over the stress test horizon), as long as these strategies depend only on the expected losses of the different groups.\\

The framework proposed in this paper can be used when the portfolio amortizes and adds new loans in a balanced way: at any time, the amortization of the previous loans is compensated for by the addition of fresh loans. More precisely, let us address the case where, for each group $g$:\\
- the initial rating profile and the rating profile of the new loans is described by the vector ${\itbf w}_g$: $w_{g,i}$ is the proportion of loans with rating $i$ in the group $g$ at time $0$, with 
$w_{g,K}=0$ and $\sum_{i=1}^{K-1} w_{g,i}=1$.\\
- a fraction $1-\kappa_g$ of loans is amortized every unit time and a fraction $\kappa_g$ of new loans with the rating profile ${\itbf w}_g$ is added every unit time.\\
- the exposure at default for the group is kept constant at ${\rm EAD}_g$ (it is of course possible to consider a time-dependent evolution).\\
This situation can be modelled in the proposed framework, provided we use the updated migration matrices
\begin{align}
{\bf M}_{g,t}^{{\itbf w}} =& (1-\kappa_g) {\bf M}_{g,t} +\kappa_g 
 {\bf M}_{g}^{\itbf w}, \\
 \quad 
 \quad
 {\bf M}_{g}^{\itbf w} =& {\bf 1} {\itbf w}_g^T = \begin{pmatrix}
 w_{g,1} & \ldots & w_{g,K-1} & 0\\
\vdots &  & \vdots & \vdots\\
 w_{g,1} & \ldots & w_{g,K-1} & 0
  \end{pmatrix} ,
  \label{def:Mgw}
\end{align}
where ${\bf 1}$ is the $K$-dimensional vector full of ones.
The expected loss is then given by (\ref{def:lossexp}-\ref{def:lossexpt}) with the matrices ${\bf M}_{g,t}^{\itbf w} $ instead of ${\bf M}_{g,t}$ and ${\rm EAD}_{g,i,t}^{\itbf w} = {\rm EAD}_g w_{g,i}$.
The unexpected loss that we address in the next section
is then given by (\ref{def:loss}-\ref{def:losst})  with the conditional matrices ${\bf M}_{g,t}^{{\itbf w}} (\bZ_t)= (1-\kappa_g) {\bf M}_{g,t} (\bZ_t) +\kappa_g 
 {\bf M}_{g}^{\itbf w}$ where ${\bf M}_{g,t} (\bZ_t) $ is given by (\ref{eq:migmat}) and $ {\bf M}_{g}^{\itbf w}$ is given by (\ref{def:Mgw}).\\

We would like now to consider the unexpected loss that is a quantile of the portfolio loss distribution. As the quantile of a portfolio loss cannot be expressed simply in terms of the quantiles of the individual losses (contrarily to the expectation), a model is needed to determine the dependence structure of the individual losses.

\section{The model for default and recovery}
\label{sec:model}
We consider a structural model such as the one proposed by Merton \cite{merton}, and considerably extended in the literature \cite{hullwhite,vasicek91,vasicek}, 
where a borrower defaults when its (normalized log) 
asset value falls below an unconditional threshold value that corresponds  to the unconditional probability of default of its group and rating. 
Default correlation is introduced by assuming that the assets of the borrowers are correlated stochastic processes.
The Basel IRB ASRF framework assumes a Gaussian copula modal with the same asset correlations between the borrowers.   
We here adopt a Gaussian copula model by following the extension with a general correlation matrix as proposed by \cite{li}.
After normalization, we can write the log asset value at time $t$ of the $q$-th borrower that belongs to the ${g}$-th group and has rating $i$ at time $t-1$ in the form \cite{pykhtin04}
\begin{align}
\label{eq:asset}
X^{(q)}_t = \ba_{{g},i,t} \cdot \bZ_t  + \sqrt{1-\ba_{{g}, i, t} \cdot {\bf C} \ba_{{g},i ,t} } \eps^{(q)}_{t}
\end{align}
where 
\begin{itemize}
\item
The random vector $\bZ_t$ contains the systematic (economic, physical, and transition) risk factors at time $t$.
The vector $\bZ_t$  is assumed to have multivariate normal distribution with mean ${\bf 0}$ and correlation matrix ${\bf C}$.
If the systematic risk factors are uncorrelated, then  they are independent and identically distributed (i.i.d.)  with standard normal distribution ${\bf C}={\bf I}$.
If the  systematic risk factors are correlated, then this general model is necessary (see Section \ref{sec:riskfactors}).
\item
The vectors $\ba_{{g},i,t}$ are the factor loadings (the correlations between the systematic risk factors and the assets) for the borrowers that belong to group $g$ and have rating $i$ at time $t-1$ (see Section \ref{sec:loading}). 
\item
The idiosyncratic factors $\eps^{(q)}_{t}$ are i.i.d. with standard normal distribution and independent from $\bZ_t$;
they model the risk specific to each borrower.
\end{itemize}

The recovery rate of the $q$-th borrower that belongs to group $g$ and has rating $i$ at time $t-1$
has the general form inspired from \cite{andersen}
\begin{align}
{\rm RR}^{(q)}_{t} &= \Phi\Big( \mu_{{g},i,t} + \sigma_{{g},i,t}\big( 
\bb_{{g},i,t} \cdot \bZ_t + \sqrt{1-\bb_{{g},i,t} \cdot {\bf C} \bb_{{g},i,t} }   \tilde{\eps}^{(q)}_{t}\big) \Big)  ,
\label{eq:lossrec}
\end{align}
where $\Phi$ is the cumulative distribution function of the standard normal distribution.
The recovery rate ${\rm RR}^{(q)}_{t}$ can be influenced by the same systematic risk factors $\bZ_t$ as the assets:
\begin{itemize}
\item
The vectors $\bb_{g,i,t}$  are the factor loadings (the correlations between the systematic risk factors and the recovery rates).
We may take $\bb_{{g},i,t}=\lambda_{{g},i,t} \ba_{{g},i,t}$ in order to simplify the model, which means that the collateral is of the same type as the principal, and then $\lambda_{{g},i,t}$ determines the dependence between the 
default occurrence and the recovery rate. 
The collateral, however, may be taken of a different type from the principal and then $\bb_{{g},i,t}$ is not collinear to $ \ba_{{g},i,t}$.
\item 
The idiosyncratic factors $\tilde{\eps}^{(q)}_{t}$ are i.i.d. with standard normal distribution and independent from $\bZ_t$ and  $\eps^{(q)}_{t}$; 
they model the risk affecting the recovery rate specific to each borrower.
\item
The parameters $\mu_{g,i,t}$ and $\sigma_{g,i,t}$ make it possible to fit observed distributions of recovery rates given default.
Note that the distribution of  ${\rm RR}^{(q)}_{t} $ given default is the distribution of  ${\rm RR}^{(q)}_{t} $ given $X^{(q)}_t$ is below the threshold value corresponding to default,
as explained in Appendix \ref{app:12}.
\item
In the simple case 
when the recovery rates are deterministic and equal to ${\rm RR}_{g,i,t}$ that depend only on the group $g$, the rating $i$, and time $t$, we have 
\begin{equation}
\mu_{{g},i,t}=\Phi^{-1}({\rm RR}_{g,i,t})
\mbox{ and }\sigma_{{g},i,t}=0,
\end{equation}
and $\bb_{g,i,t}$ plays no role (we may take $\bb_{{g},i,t}={\bf 0}$).
\item 
If $\bb_{{g},i,t} ={\bf 0}$, then the  recovery rates are random but only through the idiosyncratic risk factor.
The recovery rate ${\rm RR}^{(q)}_{t} = \Phi\big( \mu_{{g},i,t} + \sigma_{{g},i,t} 
\tilde{\eps}^{(q)}_t  \big)$ is independent from $X_t^{(q)}$ 
and the distribution of ${\rm RR}^{(q)}_{t}$ given default is of the form
$\PP( {\rm RR}^{(q)}_{t} \leq r |{\rm default}) = \Phi[ (\Phi^{-1}(r) -\mu_{{g},i,t} )/\sigma_{{g},i,t} ]$.
\item 
If $\bb_{{g},i,t} \cdot {\bf C} \bb_{{g},i,t}=1$, then the  recovery rates are random but only through the systematic risk factors. We have ${\rm RR}^{(q)}_{t} = \Phi\big( \mu_{{g},i,t} + \sigma_{{g},i,t} 
\bb_{{g},i,t} \cdot \bZ_t  \big)$ is correlated to $X_t^{(q)}$ and the distribution of ${\rm RR}^{(q)}_{t}$ given default is complex (see Appendix \ref{app:12}).
\item
 The choice of the function $\Phi$ (the cdf of the standard normal distribution) is convenient to get closed form expressions
 and it allows (with the two parameters $\mu_{g,i,t}$ and $\sigma_{g,i,t}$) to match a large diversity of recovery rate distributions.
\end{itemize}
Note that, in this random recovery model, the loss given default and the default occurrence are correlated only through the systematic risk factors.

\subsection{Unconditional migration matrices}
\label{subsec:umm}
We follow the widely adopted approach to express 
the relationship between migration matrices and the normalized log asset values \cite{belkin98}.
The rating $K$ corresponds to default, it is an absorbing state.\footnote{A specific approach could be developed when default is a non-absorbing state, as is the case for Sovereigns, as a Sovereign
might move in and out of a default rating.}
The $K\times K$ matrix ${\bf M}_{{g},t}$ has non negative entries, it satisfies $\sum_{j=1}^K ({\bf M}_{{g},t})_{ij}=1$ and $({\bf M}_{{g},t})_{KK}=1$ (see Fig.~\ref{fig:matrat}).
$({\bf M}_{{g},t})_{ij}$ is the probability for a borrower in group ${g}$ and with rating $i$ at time $t-1$ to migrate to rating $j$ at time $t$.
A borrower  in group ${g}$ with rating $i$ at time $t-1$ will migrate to a rating in the interval $[j,K]$
if its normalized log asset value falls below the unconditional threshold value $z_{{g},t,ij}$.
 The unconditional distribution of the normalized log asset value (\ref{eq:asset}) of a borrower is standard normal, 
 \begin{equation}
 \PP\big( 
X^{(q)}_{t} \leq z_{{g},t,ij} \big)
=
\Phi ( z_{{g},t,ij} ) ,
\end{equation}
 so the unconditional threshold values 
 are given in terms of quantiles of the standard normal distribution:
\begin{equation}
\label{eq:threshold}
z_{{g},t,ij} = \Phi^{-1} \Big( \sum_{j'=j}^K ({\bf M}_{{g},t})_{ij'} \Big)   .
\end{equation}
Note that:\\
- $z_{{g},t,i1}=+\infty$ for all $i \leq K$ because $\sum_{j'=1}^K ({\bf M}_{{g},t})_{ij'}=1$.\\
- The term $z_{{g},t,iK}$ is the unconditional threshold value
that corresponds  to the unconditional probability of default at time $t$ for a borrower in group ${g}$ with rating~$i$ at time $t-1$.\\
- $z_{{g},t,Kj}=+\infty$ for $j \leq K$ because $({\bf M}_{{g},t})_{KK}=1$. \\ 

\subsection{Average loss given default}
\label{subsec:lgd}
By (\ref{eq:loss})-(\ref{eq:lossrec}) the average Loss Given Default for the borrowers in group ${g}$ with rating $i$ at time $t-1$ that default at time $t$ is
\begin{equation}
{\rm LGD}_{{g},i,t}^{\rm e}=
\EE  \big[ 1- {\rm RR}_{t}^{(q)} | X_{t}^{(q)} \leq z_{{g},t,iK} \big]  ,
\end{equation}
because the event ``$X_{t}^{(q)} \leq z_{{g},t,iK}$" corresponds to default for the $q$-th borrower (which belongs to group $g$ and has  rating $i$ at time $t-1$).
As shown in Appendix \ref{app:12}, 
the average Loss Given Default for the borrowers from group ${g}$ and with rating $i$ at time $t-1$ 
who default at time $t$ depends on the rating~$i$:
\begin{align}
\label{eq:lgdhist0}
{\rm LGD}_{{g},i,t}^{\rm e}
%= \EE  \big[ 1- {\rm RR}_{{g},t}^{(q)} | X_{{g},t}^{(q)} \leq z_{{g},iK} \big]
&  = 1-
\frac{1}{({\bf M}_{{g},t})_{iK}}
\Phi_2\Big( \frac{\mu_{{g},i,t}}{\sqrt{1+\sigma_{{g},i,t}^2 }} ,
z_{{g},t,iK} ; \frac{-\rho_{{g},i,t} \sigma_{{g},i,t} }{ \sqrt{1+\sigma_{{g},i,t}^2}}\Big)  ,\\
\rho_{{g},i,t}&=  \ba_{{g},i,t}\cdot {\bf C} \bb_{{g},i,t},
\end{align}
where $\Phi_2(\cdot,\cdot;\rho)$ is the bivariate cumulative Gaussian distribution with correlation $\rho$.
As seen in Appendix \ref{app:11},
$\rho_{{g},i,t}$ is related to the Kendall rank correlation coefficient (Kendall's Tau) between the normalized log asset value $X_{t}^{(q)}$ and the recovery rate ${\rm RR}_{t}^{(q)}$ by:
\begin{align}
\label{eq:tauXR0}
\tau( X_{t}^{(q)}, {\rm RR}_{t}^{(q)}) = \frac{2}{\pi} {\rm arcsin}\big(\rho_{{g},i,t}\big).
\end{align}
Of course:
\begin{itemize}
\item
If the recovery rate and the default occurrence are independent (i.e. if $\rho_{{g},i,t}=0$), then 
$$
{\rm LGD}_{{g},i,t}^{\rm e}
  = 1-
\Phi\Big( \frac{\mu_{{g},i,t}}{\sqrt{1+\sigma_{{g},i,t}^2 }} \Big) 
$$
is equal to one minus the expected recovery rate for a borrower that belongs to group $g$ and has rating $i$ at time $t-1$:
$$
 {\rm LGD}_{{g},i,t}^{\rm e} = \EE[ 1- {\rm RR}_{t}^{(q)}] .
$$
\item
If the recovery rate is deterministic and equal to ${\rm RR}_{{g},i,t}$  (i.e. if $\sigma_{{g},i,t}=0$)
then 
\begin{equation}
 {\rm LGD}_{{g},i,t}^{\rm e} = 1- {\rm RR}_{{g},i,t} . 
\end{equation}

\end{itemize}

\section{Conditional loss}
\label{sec:condloss}
We assume that the portfolio is large.
More exactly, we assume that the portfolio contains a large number $N$ of loans without it being dominated by a few loans much larger than the rest.
This hypothesis can be formulated as the non-concentration condition 
\begin{equation}
\label{eq:ncc}
\frac{ \sum_{q=1}^N ({\rm EAD}^{(q)}_t)^2 }{\big[ \sum_{q=1}^N ({\rm EAD}^{(q)}_t)\big]^2} \stackrel{N \to \infty}{\longrightarrow}0.
\end{equation}
The left-hand side is known as the Herfindahl index, or the reciprocal of the effective number of loans \cite{roncalli}.
This hypothesis implies that the idiosyncratic risks are diversified, but not the systematic risks. 
Then the conditional loss given a trajectory  ${\bf Z}$ of the systematic (economic, physical and transition) risk factors is
\begin{align}
\label{def:loss}
L({\bf Z}) =& \sum_{t=1}^T L_t ({\bf Z}) ,\\
L_1({\bf Z}) =&   \sum_{{g}=1}^{G} \sum_{i=1}^{K-1} 
({\bf M}_{{g},1}(\bZ_1))_{iK} {\rm LGD}_{{g},i,1}(\bZ_1)
{\rm EAD}_{{g},i,1} ,
\\
L_t({\bf Z}) =& \sum_{{g}=1}^{G} \sum_{i,j=1}^{K-1} 
\big( {\bf M}_{{g},1} (\bZ_1) \cdots {\bf M}_{{g},t-1}(\bZ_{t-1})\big)_{ij} ({\bf M}_{{g},t}(\bZ_t))_{jK}
 {\rm LGD}_{{g},j,t}(\bZ_t)
 {\rm EAD}_{{g},i,t} ,
 \label{def:losst}
\end{align}
for $t\geq 2$,
where:
\begin{itemize}
\item
$L_t({\bf Z})$ is the conditional loss due to the defaults that occur at time $t$.
\item
${\bf Z}=(\bZ_1,\ldots,\bZ_T)$ is the trajectory of the systematic risk factors.
\item
$ {\rm EAD}_{{g},i,t}$ is the total exposure at default  (\ref{def:wgi}) given default at time $t$ due to the borrowers in group ${g}$ and with initial rating $i$.
\item
${\bf M}_{{g},t}(\bZ_t)$ is the conditional migration matrix (of size $K\times K$); $({\bf M}_{{g},t}(\bZ_t))_{ij}$ is the probability for a borrower in group ${g}$ and with rating $i$ at time $t-1$ to migrate to rating $j$ at time $t$, given the systematic risk factors $\bZ_t$ during this period (see Eq.~(\ref{eq:migmat})).
\item
${\rm LGD}_{{g},i,t}(\bZ_t)$ is the conditional Loss Given Default for a borrower in group ${g}$ and with rating $i$ at time $t-1$ who defaults at time $t$ (its rating jumps from $i$ to $K$), given the systematic risk factors $\bZ_t$ during this period (see Eq.~(\ref{eq:lgdcond})).
\end{itemize}

In the next subsections we present closed form expressions for the conditional migration matrix ${\bf M}_{{g},t}(\bZ_t)$ and
the conditional Loss Given Default  ${\rm LGD}_{{g},i,t}(\bZ_t)$.
As a result we have closed form expressions for the conditional loss $L({\bf Z})$ and the conditional partial losses $L_t({\bf Z})$.

Given a distribution for the process ${\bf Z}$, the conditional loss in stressed conditions $L_{\rm stress}^{1-\alpha}$ is 
the $1-\alpha$-quantile of $L({\bf Z})$:
\begin{equation}
\label{def:Lstress}
\PP( L({\bf Z}) \leq L_{\rm stress}^{1-\alpha} ) = 1-\alpha  ,
\end{equation}
with typically $\alpha=10^{-3}$ ($1-\alpha=99.9\%$) or $\alpha=10^{-3} T$ (with $T$ expressed in time units).
A straightforward method to estimate this quantile is a Monte Carlo method with a sample size $({\bf Z}^{(k)})_{k=1}^{N_{\rm MC}}$ of the order of $N_{\rm MC}=100/\alpha$. The estimator is the empirical $1-\alpha$-quantile of the sample $(L({\bf Z}^{(k)}))_{k=1}^{N_{\rm MC}}$. Variance reduction techniques,
(such as importance sampling), can be implemented to reduce the required sample size.

Given a distribution for the process ${\bf Z}$, the conditional loss in stressed conditions $L_{t,{\rm stress}}^{1-\alpha}$ during the $t$-th period (the $t$-th year when the time unit is one year), is 
the $1-\alpha$-quantile of $L_t({\bf Z})$:
\begin{equation}
\label{def:Lstresst}
\PP( L_t({\bf Z}) \leq L_{t,{\rm stress}}^{1-\alpha}) = 1-\alpha  .
\end{equation}
Note that $L_t({\bf Z})$, $t=1,\ldots,T$ are correlated 
and are, of course, correlated with $L({\bf Z})$ since $L({\bf Z})=\sum_{t=1}^T L_t({\bf Z})$.
We have:
\begin{equation}
\sum_{t=1}^T L_{t,{\rm stress}}^{1- \alpha/T} \geq L_{{\rm stress}}^{1-\alpha}  .
\end{equation}

The regulatory capital charge $K_t$ at time $t$ for the portfolio is 
\begin{equation}
K_t = L_{{\rm stress},t}^{1-\alpha} - L^{\rm e}_t,
\end{equation} 
with the expected loss $L^{\rm e}_t$ given by (\ref{def:lossexpt}) and the unexpected loss $L_{{\rm stress},t}^{1-\alpha}$ given by (\ref{def:Lstresst}). It can be multiplied by a maturity adjustment factor, given by the foundation IRB model when the unit time is one year for instance \cite{basel}.
 It could also be possible to compute an average capital charge, that would be $K/T$ where $K = L_{{\rm stress}}^{1-\alpha T} - L^{\rm e}$,
with the expected loss $L^{\rm e}$ given by (\ref{def:lossexp}) and the unexpected loss $L_{{\rm stress}}^{1-\alpha T}$ given by (\ref{def:Lstress}).

\subsection{Conditional migration matrices}
Here we assume that the unconditional migration matrices ${\bf M}_{{g},t}$ for each group ${g}$ are known and we can then express the conditional migration matrices \cite{belkin98}.
Given $\bZ_t$, a borrower  in group ${g}$ with rating $i$ at time $t-1$ will migrate to  a rating in the interval $[j,K]$ at time $t$
if its  normalized log asset value (given $\bZ_t$) falls below  the threshold $z_{{g},t,ij}$. This event has probability
\begin{equation}
\PP\Big( 
X^{(q)}_{t} \leq z_{{g},t,ij} | \bZ_t 
\Big)
=
\Phi \Big( \frac{ z_{{g},t,ij} -  \ba_{{g},i,t} \cdot  \bZ_t }{\sqrt{1-\ba_{{g},i,t} \cdot {\bf C} \ba_{{g},i,t}}}\Big)   .
\end{equation}
As a consequence, the conditional migration matrix ${\bf M}_{{g},t}(\bZ_t)$ is given by
\begin{align}
({\bf M}_{{g},t}(\bZ_t))_{ij}
=
\left\{
\begin{array}{ll}
\displaystyle
1 -\Phi  \Big( \frac{ z_{{g},t,i2} -  \ba_{{g},i,t} \cdot \bZ_t }{ \sqrt{1-\ba_{{g},i,t} \cdot {\bf C} \ba_{{g},i,t} } }\Big),  & \mbox{ if } j=1,\\
\displaystyle
\Phi  \Big( \frac{ z_{{g},t,ij} -  \ba_{{g},i,t} \cdot \bZ_t }{ \sqrt{1-\ba_{{g},i,t} \cdot {\bf C} \ba_{{g},i,t} } }\Big)  \\
\hspace*{0.25in} 
\displaystyle
 -\Phi  \Big( \frac{ z_{{g},t,ij+1} -  \ba_{{g},i,t}\cdot \bZ_t }{ \sqrt{1-\ba_{{g},i,t} \cdot {\bf C} \ba_{{g},i,t} } }\Big) , & \mbox{ if } 2\leq j \leq K-1,\\
\displaystyle
\Phi  \Big( \frac{ z_{{g},t,iK} -  \ba_{{g},i,t} \cdot \bZ_t }{ \sqrt{1-\ba_{{g},i,t} \cdot {\bf C} \ba_{{g},i,t} } }\Big) ,  & \mbox{ if } j=K.
\end{array}
\right.
\label{eq:migmat}
\end{align}
The entries of the last column $({\bf M}_{{g},t}(\bZ_t))_{iK}$ gives the conditional probability of default for a borrower in group ${g}$ with rating $i =1,\ldots,K-1$ at time $t-1$.

\subsection{Conditional loss given default}
The particular form of the loss (\ref{eq:loss})-(\ref{eq:lossrec}) makes it possible to give a simple closed form formula for the 
 conditional Loss Given Default ${\rm LGD}_{{g},i,t}(\bZ_t)$:
\begin{equation}
{\rm LGD}_{{g},i,t}(\bZ_t) = 1 -\Phi \Big( \frac{\mu_{{g},i,t} + \sigma_{{g},i,t} \bb_{{g},i,t}\cdot\bZ_t }{\sqrt{1+\sigma^2_{{g},i,t}(1-\bb_{{g},i,t}\cdot {\bf C} \bb_{{g},i,t}})}\Big)   .
\label{eq:lgdcond}
\end{equation}
Of course, if the recovery rate is deterministic and equal to ${\rm RR}_{{g},i,t}$  (i.e. if $\sigma_{{g},i,t}=0$)
then 
\begin{equation}
{\rm LGD}_{{g},i,t}(\bZ_t) = 1- {\rm RR}_{{g},i,t} . \end{equation}

\subsection{An explicit and simple case}
If $T=1$, ${G}=1$, ${\bf C}={\bf I}$, $\sigma_{1,i,1}=0$ (the LGD is deterministic), 
and $\ba_{1,i,1}=\ba_1$ (all borrowers have the same exposition with respect to the systematic risks whatever their inital rating),  
then we can get a closed form expression for the conditional loss in stressed conditions $L_{\rm stress}^{1-\alpha}$.
Indeed we have
\begin{align}
\nonumber
L({\bf Z}) &=
\sum_{i=1}^{K-1} 
({\bf M}_{1,1}(\bZ_1))_{iK} (1-{\rm RR}_{1,i,1}) {\rm EAD}_{1,i,1} \\
& =  {\cal L}( \ba_{1} \cdot \bZ_1 ) ,
\end{align}
where
\begin{align}
{\cal L}(z)& = 
\sum_{i=1}^{K-1}
\Phi  \Big( \frac{ z_{1,1,iK} -  z }{\sqrt{1-\|\ba_{1}\|^2}}\Big) 
(1-{\rm RR}_{1,i,1})  {\rm EAD}_{1,i,1}.
\end{align}
The function $z \mapsto {\cal L}(z)$ is decreasing, so we have for any $\ell$:
\begin{align}
\nonumber
\PP\big( L({\bf Z})  \leq \ell \big) &= \PP \big( \ba_{1}\cdot \bZ_1 \geq {\cal L}^{-1}(\ell)  \big)\\ 
&=1 -\Phi\Big(  \frac{{\cal L}^{-1}(\ell)}{\|\ba_{1}\|}  \Big) = \Phi\Big(  -\frac{{\cal L}^{-1}(\ell)}{\|\ba_{1}\|}  \Big)    ,
\end{align}
because the random variable $\ba_{1}\cdot\bZ_1$,
has distribution ${\cal N}(0,\|\ba_{1}\|^2)$.
The conditional loss  in stressed conditions (\ref{def:Lstress}) is therefore with $\alpha=10^{-3}$
(as in \cite{roncalli}):
\begin{align}
\nonumber
L_{\rm stress}^{0.999} &= {\cal L}\big( - \Phi^{-1}(0.999) \|\ba_{1}\| \big)\\
\nonumber
&=\sum_{i=1}^{K-1} 
\Phi  \Big( \frac{ z_{1,1,iK} +\Phi^{-1}(0.999) \|\ba_{1}\| }{\sqrt{1-\|\ba_{1}\|^2}}\Big)(1-{\rm RR}_{1,i,1})   {\rm EAD}_{1,i,1}.
\end{align}

\section{Model for the systematic risk factors}
\label{sec:riskfactors}
The vector $\bZ_t$ contains $d$ systematic risk factors.

\subsection{Independent risk factors}
Here we consider models in which $\bZ_t$ has i.i.d. entries with standard normal distribution, i.e.  ${\bf C}={\bf I}$.

When one wishes to study economic systematic risk, one usually uses a one-factor model. In this model the $Z_t$ are i.i.d. with standard normal distribution.

Here we want to study economic, physical, and transition systematic risks.
One can therefore consider a three-factor model $\bZ_t=(Z_{t,1},Z_{t,2},Z_{t,3})$ where $Z_{t,1}$ is the economic risk factor,
$Z_{t,2}$ is the transition risk factor, and $Z_{t,3}$ is the physical risk factor.
We can take them independently with standard normal distribution.

We can also make the model more complex by considering several independent physical risk factors, one per geographical region.
If we want to model groups that are exposed to only one regional physical risk, then we would need to index the group as follows: ${g}=(e,r)$, where $e=1,\ldots,E$ is the index of the non-geographical sector (economic sector and/or climate risk mitigation and adaptation strategy and/or collateral type),
and
$r=1,\ldots,R$ the index of the geographical region, so that there are $G=ER$ groups in total.
The vector $\bZ_t$ would then be of the form $\bZ_t=(Z_{t,j})_{j=1}^{2+R}$ where $Z_{t,1}$ is the economic risk factor,
$Z_{t,2}$ is the transition risk factor, and $Z_{t,(2+r)}$ is the physical risk factor of the $r$-th region, $r=1,\ldots,R$.
The factor loadings would then be of the form $\ba_{e,r,i,t} = (a_{e,r,i,t,j})_{j=1}^{2+R} $, where $a_{e,r,i,t,1}$ is the factor loading associated to the economic risk at time $t$ of a borrower with rating $i$ in non-geographical sector $e$ and region $r$, $a_{e,r,i,t,2}$ is the factor loading associated to the transition risk for such a borrower,
$a_{e,r,i,t,2+r}$ is the factor loading associated to the physical risk of the $r$-th region for such a borrower, the factor loadings associated to the physical risks of the other regions are zero:
$a_{e,r,i,t,2+r'}=0$  for $r'\neq r$. 
We can, for instance, also introduce other groups that are exposed to several regional physical risks simultaneously.

\subsection{Correlated risk factors}
\label{subsec:corrfac}
Here we consider models in which $\bZ_t$ has multivariate normal distribution with mean ${\bf 0}$ and correlation matrix ${\bf C}$.
These models are necessary if we want to model
correlations between some systematic risk factors.

We may think at an example where $\bZ_t=(Z_{t,j})_{j=1}^{2+R}$, $Z_{t,1}$ is the economic risk factor,
$Z_{t,2}$ is the transition risk factor, and $Z_{t,2+r}$ is the physical risk factor of the $r$-th region, $r=1,\ldots,R$:
\begin{equation}
\bZ_{t} \sim {\cal N}({\bf 0} , {\bf C}) \mbox{ i.i.d.}  ,
\end{equation}
with
\begin{equation}
{\bf C} = \begin{pmatrix}
1 & -\rho & 0 & 0 & \cdots &0\\
-\rho &1 & 0 &0 & \cdots &0\\
0 & 0 & 1& \rho_o &\cdots &  \rho_o \\
0 & 0&  \rho_o & 1 &\cdots &  \rho_o \\
0 & 0&  \rho_o &  \rho_o & \ddots &  \rho_o \\
0& 0&  \rho_o &  \rho_o & \rho_o & 1
\end{pmatrix}  ,
\label{eq:modelmatC}
\end{equation}
which means that:\\
1) the physical risks of different geographical regions are positively correlated ($\rho_o \in (0,1)$)
and independent from the economic and transition risks, \\
2) the transition risk is negatively correlated with the economic risk $(\rho \in (0,1)$). This comes from the observation that an economic downturn may involve a reduction in emissions of greenhouse gases.\\
The covariance matrix (\ref{eq:modelmatC}) can be made more complex, for instance, if correlations between physical risks in different regions are known (based on distances for instance).

\section{Model for the loading factors}
\label{sec:loading}

\subsection{The model with a unique systematic risk factor}
Under the foundation
 IRB (Internal Rating Based) approach \cite{basel}:\\
- The time unit is one year.\\
- The LGD model is deterministic 
and imposed by the regulator. That is to say, ${\rm LGD}_{g,i,t}$ are given by ${\rm LGD}_{g,i}^{\rm reg}$ that do not depend on $t$, but that depend on the group $g$ and the rating $i$ before default.\\
- The EAD model is deterministic and determined by the loan composition of the portfolio.\\
- The unconditional migration matrices ${\bf M}_{g,t}$ are given by ${\bf M}_{g}^{\rm reg}$ that do not depend on $t$, but depend on the group $g$. The matrices ${\bf M}_{g}^{\rm reg}$ are typically estimated from historical data and provided by rating agencies.\\
- The correlation model to economic risk (assumed to be the unique systematic risk factor) is determined by a formula that is imposed by the regulator and that is a function of the probability of default \cite{basel,basel3}:
\begin{align}
R_{g,i}^{\rm reg} =& {\cal R}({\rm PD}_{g,i}^{\rm reg}),\\
{\cal R}({\rm PD})  =& 0.12 \frac{1-e^{- 50 {\rm PD}} }{1-e^{-50}}
+0.24 \Big(1-  \frac{1-e^{- 50 {\rm PD}} }{1-e^{-50}}\Big)   ,
\label{eq:formreg}
\end{align}
where 
${\rm PD}_{g,i}^{\rm reg}= ({\bf M}_{g}^{\rm reg})_{iK}$ is the probability of default at time $t$ of a borrower in group $g$ and with rating $i$ at time $t-1$.\footnote{The formula (\ref{eq:formreg}) is proposed in the  Basel II IRB approach \cite{basel}. It was updated in the Basel III IRB approach and multiplied by $1.25$ \cite{basel3}.}
There is a unique systematic risk factor $Z_{t,1}$ and the  loading factor does not depend on $t$ and is equal to
 $a_{g,i,t,1} = a_{g,i}^{\rm reg}$, with
 \begin{equation}
a_{g,i}^{\rm reg}=
\sqrt{ R_{g,i}^{\rm reg} }.
\end{equation}
Under these hypotheses, the expected loss is given by (\ref{def:lossexp}):
\begin{align*}
L^{\rm e} =& \sum_{t=1}^T L_t^{\rm e} ,\\
L_1^{\rm e} =&
  \sum_{{g}=1}^{G} \sum_{i=1}^{K-1} 
({\bf M}_{{g}}^{\rm reg})_{iK} {\rm LGD}_{{g},i}^{\rm reg}  {\rm EAD}_{{g},i,1} ,
\\
L_t^{\rm e} =& \sum_{{g}=1}^{G} \sum_{i,j=1}^{K-1} 
\big( ({\bf M}_{{g}}^{\rm reg}  )^{t-1}\big)_{ij} \big({\bf M}_{{g}}^{\rm reg} \big)_{jK}
 {\rm LGD}_{{g},j}^{\rm reg} {\rm EAD}_{{g},i,t}  ,
\end{align*}
for $t\geq 2$, 
and the conditional loss given a trajectory  ${\bf Z}=(Z_{1,1},\ldots,Z_{T,1})$ of the economic risk factor is given by (\ref{def:loss}):
\begin{align*}
L({\bf Z}) =& \sum_{t=1}^T L_t({\bf Z}) ,\\
L_1({\bf Z}) =&
  \sum_{{g}=1}^{G} \sum_{i=1}^{K-1} 
({\bf M}_{{g}}^{\rm reg}(Z_{1,1}))_{iK} {\rm LGD}_{{g},i}^{\rm reg}  {\rm EAD}_{{g},i,1} ,
\\
L_t({\bf Z}) =& \sum_{{g}=1}^{G} \sum_{i,j=1}^{K-1} 
\big( {\bf M}_{{g}}^{\rm reg} (Z_{1,1}) \cdots {\bf M}_{{g}}^{\rm reg}(Z_{t-1,1})\big)_{ij} \big({\bf M}_{{g}}^{\rm reg}(Z_{t,1}) \big)_{jK}
 {\rm LGD}_{{g},j}^{\rm reg} {\rm EAD}_{{g},i,t}  ,
\end{align*}
for $t\geq 2$, where the conditional migration matrices are given by (\ref{eq:migmat}).
The conditional loss  in stressed conditions $L_{\rm stress}^{1-\alpha}$ is 
the $1-\alpha$-quantile of $L({\bf Z})$ when the $Z_{t,1}$ are independent and identically distributed with the standard normal distribution.\\

\subsection{The model with multiple systematic risk factors}

We need to extend the previous model to take into account transition and physical risks.
We still assume that $ {\rm LGD}_{g,i}^{\rm reg}$  and $ {\bf M}_{g}^{\rm reg}$  are given. We need to extend the correlation model and its relation to the loading factors~$\ba_{g,i,t}$
for the systematic risk factors $\bZ_t = (Z_{t,j})_{j=1}^{R+2}$ described in Section~\ref{subsec:corrfac}.

We introduce the macro-correlation parameters 
${\bzeta}_t = ({\zeta}_{t,j})_{j=1}^{R+2}$. They give the evolution of the intensities of the $R+2$ systematic risk factors (economic, transition, physical divided into $R$ regions). $\zeta_{t,1}$ is associated to the economic risk and assumed to be constant and equal to $\zeta_1$. $\zeta_{t,2}$ and $\zeta_{t,2+r}$ are associated to the transition and physical risks and evolve in time. These parameters are relative to each other and should be expressed in the same ``units". For instance, we may express all macro-correlation parameters in terms of GDP growth rates. 
$\zeta_{1}$ can be the GDP growth rate involved by an economic downturn.
$\zeta_{t,2}$ can be calibrated from the Intergovernmental Panel on Climate Change (IPCC) carbon emission pathway expressed in impact to GDP growth rate. $\zeta_{t,2+r}$ can be calibrated from the IPCC GDP growth rate assessment for the region~$r$.
Macro-economic and macro-climatic data can also be obtained from the Network for Greening the Financial System (NGFS) or the International Energy Agency (IEA).
From now on we assume that ${\bzeta}_t$ is given.

We introduce the micro-correlation adjustment parameters $\alpha_{g,i,t,j}$.
Each borrower in group $g$ and with rating $i$ at time $t-1$ has a micro-correlation adjustment parameter $\alpha_{g,i,t,j}$ to the $j$-th systematic risk factor.
This micro-correlation parameter depends on the group. It may depend on the rating. It may be time-dependent in order to take into account mitigation and adaptation efforts by the borrowers.
Note that a micro-correlation adjustment parameter can be negative (for instance, transition risk may favour a green economic sector).
From now on we assume that ${\balpha}_{g,i,t} = (\alpha_{g,i,t,j})_{j=1}^{R+2}$ are given.

We introduce the correlation $R_{g,i,t}$.
The correlation is the proportion of the variance of the  normalized  log asset value that is due to the systematic risks. Equivalently, $1-R_{g,i,t}$ is the proportion of  the variance of the normalized log asset value that is due to the idiosyncratic risk of a borrower. From (\ref{eq:asset}) it is related to the factor loadings through the relation:
\begin{equation}
R_{g,i,t} = \ba_{g,i,t}\cdot {\bf C} \ba_{g,i,t} .
\end{equation}

The approach that we propose is described in subsection \ref{subsubsec:T3}. We first describe two tentative approaches that turned out to be inconsistent and that motivate the approach proposed in subsection \ref{subsubsec:T3}.

\subsubsection{\it First tentative approach (approach T1) for the correlation model and factor loadings.}
We consider here that:\\
- the time unit is one year,\\
- the migration matrices ${\bf M}_{g,t}$ do not depend on $t$ and are equal to ${\bf M}^{\rm reg}_g$,\\
- the correlation $R_{g,i,t}$ at any time $t$ is determined by the regulator's formula which does not depend on $t$,\\
- the factor loadings $a_{g,i,t,j}$ are proportional to the product of the macro-correlation and micro-correlation adjustment parameters.

As a result, we have
\begin{equation}
R_{g,i,t} = R_{g,i}^{\rm reg} ,\quad \quad R_{g,i}^{\rm reg}= {\cal R}({\rm PD}_{g,i}^{\rm reg}) ,
\end{equation}
with ${\rm PD}_{g,i}^{\rm reg}= ({\bf M}_{g}^{\rm reg})_{iK}$,  ${\cal R}$ defined by (\ref{eq:formreg}), and
\begin{equation}
\label{eq:agit:approach1}
a_{g,i,t,j} = \sqrt{R_{g,i}^{\rm reg}} \frac{\tilde{a}_{g,i,t,j}}{\sqrt{\tilde{\ba}_{g,i,t} \cdot {\bf C} \tilde{\ba}_{g,i,t}}}  ,
\end{equation}
with
\begin{equation}
\tilde{a}_{g,i,t,j} = 
\alpha_{g,i,t,j} 
\zeta_{t,j} .
\end{equation}

\noindent
{\it Proof.}
The factor loadings $a_{g,i,t,j}$ are proportional to the product $\tilde{a}_{g,i,t,j}$ of the macro-correlation and micro-correlation adjustment parameters.
From (\ref{eq:asset}) 
the factor loadings also satisfy $\ba_{g,i,t}\cdot {\bf C} \ba_{g,i,t}=R_{g,i}^{\rm reg}$.
This imposes the form (\ref{eq:agit:approach1}) of the factor loadings.
\qed

{\bf Discussion.}
In the approach T1, when the intensities $\zeta_{t,2}$ and/or $\zeta_{t,2+r}$ increase (compared to $\zeta_1$ that is constant), then the correlation $R_{g,i,t}$ is not affected because it is determined by the regulator's formula, which depends only on the given unconditional migration matrices.
The only effect of the increase of the intensities $\zeta_{t,2}$ and/or $\zeta_{t,2+r}$ is to modify the proportions of the economic and climate contributions to the constrained value of the correlation $R^{\rm reg}_{{g},i}$.
To sum-up, if the climatic risk intensities increase, then the economic risk intensity decays  in order to maintain the correlation value $ R_{g,i}^{\rm reg}$. 
This makes the approach T1 not appropriate.

\subsubsection{\it Second tentative approach (approach T2) for the correlation model and loading factors.}
We consider here that:\\
- the time unit is one year,\\
- the migration matrices ${\bf M}_{g,t}$ do not depend on $t$ and are equal to ${\bf M}^{\rm reg}_g$,\\
- the correlation $R_{g,i,1}$ at time $1$ is determined by the regulator's formula, but this formula is updated at time $t\geq 2$ because, contrary to the economic risk, which is stationary, the physical and transition risks evolve in time.\\
- the factor loadings $a_{g,i,t,j}$ are proportional to the product of the macro-correlation and micro-correlation adjustment parameters.

As a result we have
\begin{equation}
{R}_{g,i,t} = 
 \frac{
 \tilde{\ba}_{g,i,t} \cdot {\bf C} \tilde{\ba}_{g,i,t}  {R}_{g,i}^{\rm reg} }
{\tilde{\ba}_{g,i,t} \cdot {\bf C} \tilde{\ba}_{g,i,t} {R}_{g,i}^{\rm reg}
+
\tilde{\ba}_{g,i,1} \cdot {\bf C} \tilde{\ba}_{g,i,1}(1-  {R}_{g,i}^{\rm reg}) }  ,
\label{eq:expresscorrapp2}
\end{equation}
and
\begin{equation}
a_{g,i,t,j} =
 \sqrt{
 {R}_{g,i}^{\rm reg}
 }
 \frac{
 \tilde{a}_{g,i,t,j}   
}
{
\sqrt{
\tilde{\ba}_{g,i,t} \cdot {\bf C} \tilde{\ba}_{g,i,t} {R}_{g,i}^{\rm reg}
+
\tilde{\ba}_{g,i,1} \cdot {\bf C} \tilde{\ba}_{g,i,1}(1-  {R}_{g,i}^{\rm reg}) 
}
}
,
\label{eq:agit:approach2}
\end{equation}
with
\begin{equation}
\tilde{a}_{g,i,t,j} = 
\alpha_{g,i,t,j} 
\zeta_{t,j} .
\end{equation}

{\it Proof.}
At time $1$ (see the approach T1) the normalized log asset value is given by 
\begin{equation}
\label{eq:assetreg1}
X^{(q)}_1 = \ba_{{g},i}^{\rm reg} \cdot \bZ_1 + \sqrt{1-\ba_{{g}, i}^{\rm reg} \cdot {\bf C} \ba_{{g},i}^{\rm reg} } \eps^{(q)}_{1},
\end{equation}
with 
$$
a^{\rm reg}_{g,i,j} = \frac{ \sqrt{R_{g,i}^{\rm reg}} \tilde{a}_{g,i,1,j}}{\sqrt{\tilde{\ba}_{g,i,1} \cdot {\bf C} \tilde{\ba}_{g,i,1}}}  ,
\quad \quad 
\tilde{a}_{g,i,1,j} = 
\alpha_{g,i,1,j} 
\zeta_{1,j} ,\quad \quad R_{g,i}^{\rm reg}= {\cal R}(({\bf M}_{g}^{\rm reg})_{iK}) .
$$
If the loading factors were stationary (time-independent), we would have for any time $t$
$$
X^{(q)}_t = \ba_{{g},i}^{\rm reg} \cdot \bZ_t + \sqrt{1-\ba_{{g}, i}^{\rm reg} \cdot {\bf C} \ba_{{g},i}^{\rm reg} } \eps^{(q)}_{t},
$$
and the approach T1 would be valid.
However, the micro-correlation and macro-correlation parameters evolve in time so we need to update this representation.\\
The unnormalized log asset value is given by 
$$
\tilde{X}^{(q)}_t = \tilde{\ba}_{{g},i,t} \cdot \bZ_t + 
\tilde{\sigma}_{{g},i} \eps^{(q)}_{t},
$$
which is a Gaussian variable with mean zero and variance $\tilde{\ba}_{{g}, i,t} \cdot {\bf C} \tilde{\ba}_{{g},i,t}  + 
\tilde{\sigma}^2_{{g},i}$.
Here $\tilde{a}_{g,i,t,j} = 
\alpha_{g,i,t,j} 
\zeta_{t,j} $ and $\tilde{\sigma}_{{g},i}$ does not depend on $t$ because the unnormalized idiosyncratic risk is assumed to be stationary.\\
The normalized log asset value $X^{(q)}_1$ needs to be of variance one so that the migration matrix ${\bf M}_{{g},1}$ is equal to ${\bf M}^{\rm reg}_{g}$.
This means that ${X}^{(q)}_1 = \tilde{X}^{(q)}_1 / 
\sqrt{\tilde{\ba}_{{g}, i,1} \cdot {\bf C} \tilde{\ba}_{{g},i,1} + 
\tilde{\sigma}^2_{{g},i}}$.
Since ${X}^{(q)}_1$ is of the form (\ref{eq:assetreg1}), the variance $\tilde{\sigma}_{{g},i}^2$ solves
$$
1-\ba_{{g}, i}^{\rm reg} \cdot {\bf C} \ba_{{g},i}^{\rm reg}  
=
\frac{ 
\tilde{\sigma}_{{g},i}^2 }{
\tilde{\ba}_{{g}, i,1} \cdot {\bf C} \tilde{\ba}_{{g},i,1} + 
\tilde{\sigma}^2_{{g},i} } ,
$$
which gives, with the identity $R_{g,i}^{\rm reg} = \ba^{\rm reg}_{g,i} \cdot {\bf C} \ba^{\rm reg}_{g,i}$,
$$
\tilde{\sigma}_{{g},i}^2
=
 \tilde{\ba}_{g,i,1} \cdot {\bf C} \tilde{\ba}_{g,i,1} \frac{1-{R}_{g,i}^{\rm reg} }{{R}_{g,i}^{\rm reg} }  .
$$
The normalized log asset value $X^{(q)}_t$ needs to be of variance one so that the migration matrix ${\bf M}_{{g},t}$ is equal to ${\bf M}^{\rm reg}_{g}$.
This means that ${X}^{(q)}_t = \tilde{X}^{(q)}_t / 
\sqrt{\tilde{\ba}_{{g}, i,t} \cdot {\bf C} \tilde{\ba}_{{g},i,t} + 
\tilde{\sigma}^2_{{g},i}} $. Since  ${X}^{(q)}_t$ is of the form (\ref{eq:asset}),
we find that  the factor loadings $a_{g,i,t,j}$ are of the form
$$
a_{g,i,t,j}  =  \frac{\tilde{a}_{g,i,t,j}}{\sqrt{\tilde{\ba}_{{g}, i,t} \cdot {\bf C} \tilde{\ba}_{{g},i,t} + 
\tilde{\sigma}^2_{{g},i}}  } ,
$$
which gives (\ref{eq:agit:approach2}),
and the correlation is of the form
$$
{R}_{g,i,t} = \frac{\tilde{\ba}_{g,i,t} \cdot {\bf C} \tilde{\ba}_{g,i,t}}{\tilde{\ba}_{g,i,t} \cdot {\bf C} \tilde{\ba}_{g,i,t} +\tilde{\sigma}^2_{g,i}}  .
$$
which gives (\ref{eq:expresscorrapp2}).
\qed

{\bf Discussion.}
In the approach T2, the correlation $R_{g,i,t}$ is different from 
(typically, larger than)
 ${R}_{g,i}^{\rm reg}$, which means that the exposition to the systematic risk factors is different from
  (typically larger than) 
 the exposition defined by the regulator. As the migration matrices are assumed to be constant and given by ${\bf M}^{\rm reg}_{g}$, this means that the exposition to the idiosyncratic risks $\sqrt{1-R_{g,i,t}^2}$ is different (typically, smaller than) the exposition defined by the regulator. 
To sum-up, if the climatic risk intensities increase, then the idiosyncratic risk decays in order to maintain the same unconditional migration matrices.
This makes the approach T2 not appropriate.

\subsubsection{\it Proposed approach for the correlation model and loading factors.}
\label{subsubsec:T3}%
We consider here that:\\
- the time unit is one year,\\
- at time $1$ the migration matrix ${\bf M}_{g,1}$ is equal to ${\bf M}^{\rm reg}_g$ and the correlation $R_{g,i,1}$  is determined by the regulator's formula,\\
- the migration matrices and the regulator's formula for the correlation are updated at time $t\geq 2$ because, contrary to the economic and idiosyncratic risks, 
 which are stationary, the physical and transition risks evolve in time.\\
- the factor loadings $a_{g,i,t,j}$ are proportional to the product of the macro-correlation and micro-correlation adjustment parameters.

As a result,
at time $1$, the formulas are reduced to the formulas of the approach T1:
\begin{align}
{\bf M}_{g,1} =&
{\bf M}_{g}^{\rm reg} , \\
R_{g,i,1} =& R_{g,i}^{\rm reg} ,
\quad \quad R_{g,i}^{\rm reg}= {\cal R}(({\bf M}_{g}^{\rm reg})_{iK}) ,
\\
a_{g,i,1,j} =& a_{g,i,j}^{\rm reg},
\quad \quad
a_{g,i,j}^{\rm reg}=
\sqrt{R_{g,i}^{\rm reg}} \frac{\tilde{a}_{g,i,1,j}}{\sqrt{\tilde{\ba}_{g,i,1} \cdot {\bf C} \tilde{\ba}_{g,i,1}}}  ,
\end{align}
with ${\cal R}$ defined by (\ref{eq:formreg}) and
\begin{equation}
\tilde{a}_{g,i,t,j} = 
\alpha_{g,i,t,j} 
\zeta_{t,j} .
\end{equation}
At time $t\geq 1$, 
we have
\begin{equation}
\label{eq:defM:approach3}
({\bf M}_{g,t})_{ij} =
\left\{
\begin{array}{ll}
1- \Phi ( z_{g,t,i2}) & \mbox{ if } j=1,\\ 
 \Phi ( z_{g,t,ij}) - \Phi( z_{g,t,ij+1})  & \mbox{ if } 2\leq j \leq K-1,\\
 \Phi ( z_{g,t,iK}) & \mbox{ if } j=K,
 \end{array}
 \right.
\end{equation}
with
\begin{align}
z_{g,t,ij}=& \frac{z_{g,ij}^{\rm reg}}{
\sqrt{
1+ {\bc}_{g,i,t} \cdot {\bf C} {\bc}_{g,i,t} - 
\ba_{g,i}^{\rm reg}\cdot {\bf C} \ba_{g,i}^{\rm reg}}
} ,
\\
z_{g,ij}^{\rm reg}=& \Phi^{-1} \Big( \sum_{j'=j}^K ({\bf M}_g^{\rm reg})_{ij'}\Big) ,\\
{c}_{g,i,t,j} =& a_{g,i,j}^{\rm reg} \frac{\tilde{a}_{g,i,t,j}}{\tilde{a}_{g,i,1,j}}, 
\label{def:cg}
\end{align}
and we have
\begin{align}
\label{eq:defR:approach3}
R_{g,i,t} = &
\frac{ {\bc}_{g,i,t} \cdot {\bf C} {\bc}_{g,i,t}  }{
1+ {\bc}_{g,i,t} \cdot {\bf C} {\bc}_{g,i,t} - 
\ba_{g,i}^{\rm reg}\cdot {\bf C} \ba_{g,i}^{\rm reg}
} ,\\
\label{eq:defa:approach3}
a_{g,i,t,j} = &
\frac{ {c}_{g,i,t,j}}{
\sqrt{
1+ {\bc}_{g,i,t} \cdot {\bf C} {\bc}_{g,i,t} - 
\ba_{g,i}^{\rm reg}\cdot {\bf C} \ba_{g,i}^{\rm reg}}
} .
\end{align}
Note that the formulas (\ref{eq:defa:approach3}) and (\ref{eq:agit:approach2}) for the loading factors coincide, and the formulas (\ref{eq:defR:approach3}) and (\ref{eq:expresscorrapp2}) for the correlations coincide.
The difference between this approach and the approach T2 is that the migration matrices are constant in the approach T2 (which makes the approach not consistent as discussed above) while they evolve in a consistent way in this approach.

\noindent
{\it Proof.}
At time $1$  (see the approach T1) the normalized log asset value is given by 
$$
X^{(q)}_1 = \ba_{{g},i}^{\rm reg} \cdot \bZ_1 + \sqrt{1-\ba_{{g}, i}^{\rm reg} \cdot {\bf C} \ba_{{g},i}^{\rm reg} } \eps^{(q)}_{1}.
$$
If the loading factors were stationary (time-independent), we would have for any time $t$
$$
X^{(q)}_t = \ba_{{g},i}^{\rm reg} \cdot \bZ_t + \sqrt{1-\ba_{{g}, i}^{\rm reg} \cdot {\bf C} \ba_{{g},i}^{\rm reg} } \eps^{(q)}_{t},
$$
and the approach T1 would be valid.
However, the idiosyncratic risk is stationary but the micro-correlation and macro-correlation parameters evolve in time. 
This means that, using (\ref{def:cg}),
we have in fact
$$
\overline{X}^{(q)}_t = \bc_{{g},i,t} \cdot \bZ_t + \sqrt{1-\ba_{{g}, i}^{\rm reg} \cdot {\bf C} \ba_{{g},i}^{\rm reg} } \eps^{(q)}_{t},
$$
which is a Gaussian variable with mean zero and variance $1-\ba_{{g}, i}^{\rm reg} \cdot {\bf C} \ba_{{g},i}^{\rm reg}  + 
\bc_{{g},i,t}\cdot {\bf C} \bc_{{g},i,t}$.
As a consequence, the probabilities of rating change are
$$
({\bf M}_{g,t})_{ij} = 
\PP\big( \overline{X}^{(q)}_t \in [ z_{g,ij+1}^{\rm reg} , z_{g,ij}^{\rm reg}]\big) ,
$$
where the $z_{g,ij}^{\rm reg}$'s are the threshold values associated to the given unconditional migration matrix
${\bf M}_{g}^{\rm reg}$.
This gives (\ref{eq:defM:approach3}).
Furthermore, after normalization, the log asset value 
$X_t^{(q)} = \overline{X}^{(q)}_t / \sqrt{1-\ba_{{g}, i}^{\rm reg} \cdot {\bf C} \ba_{{g},i}^{\rm reg}  + 
\bc_{{g},i,t}\cdot {\bf C} \bc_{{g},i,t}}$
has now the form
$$
X^{(q)}_t = \ba_{{g},i,t} \cdot \bZ_t + \sqrt{1-\ba_{{g}, i,t} \cdot {\bf C} \ba_{{g},i,t} } \eps^{(q)}_{t},
$$
with $\ba_{{g},i,t}$ given by (\ref{eq:defa:approach3}), which also gives  (\ref{eq:defR:approach3}).
\qed

{\bf Discussion.}
In this approach, if the climatic (physical and/or transition) risk intensities increase, then the idiosyncratic risk and the economic risk stay constant, so that the overall risk increases and the
unconditional migration matrices change.
These changes are evaluated automatically from the climate scenario.

Under these hypotheses, the expected loss is given by (\ref{def:lossexp}):
\begin{align}
L^{\rm e} =& \sum_{t=1}^T L_t^{\rm e} ,\\
L_1^{\rm e} =&
  \sum_{{g}=1}^{G} \sum_{i=1}^{K-1} 
({\bf M}_{{g},1})_{iK} {\rm LGD}_{{g},i}^{\rm reg}  {\rm EAD}_{{g},i,1} ,
\\
L_t^{\rm e} =& \sum_{{g}=1}^{G} \sum_{i,j=1}^{K-1} 
\big( {\bf M}_{{g},1} \cdots {\bf M}_{{g},t-1}  \big)_{ij} \big({\bf M}_{{g},t} \big)_{jK}
 {\rm LGD}_{{g},j}^{\rm reg} {\rm EAD}_{{g},i,t}  ,
\end{align}
for $t\geq 2$,
and 
the conditional loss given a trajectory  ${\bf Z}= (\bZ_1,\ldots,\bZ_T)$ of the systematic risk factors is given by (\ref{def:loss}):
\begin{align}
L({\bf Z}) =& \sum_{t=1}^T L_t({\bf Z}) ,
\\
L_1({\bf Z}) =& \sum_{{g}=1}^{G} \sum_{i=1}^{K-1}
({\bf M}_{{g},1}(\bZ_1))_{iK} {\rm LGD}_{{g},i}^{\rm reg} 
{\rm EAD}_{{g},j,1} ,
\\
L_t({\bf Z}) =&
\sum_{{g}=1}^{G} \sum_{i,j=1}^{K-1} 
\big( {\bf M}_{{g},1} (\bZ_1) \cdots {\bf M}_{{g},t-1} (\bZ_{t-1})  \big)_{ij} \big({\bf M}_{{g},t} (\bZ_t) \big)_{jK}
 {\rm LGD}_{{g},j}^{\rm reg}{\rm EAD}_{{g},i,t}  ,
\end{align}
for $t \geq 2$, where the conditional migration matrices are given by (\ref{eq:migmat}).
The conditional loss  in stressed conditions $L_{\rm stress}^{1-\alpha}$ is 
the $1-\alpha$-quantile of $L({\bf Z})$ when the $\bZ_{t}$ are independent and identically distributed with the multivariate normal distribution ${\cal N}({\bf 0}, {\bf C})$ (with $ {\bf C}$ given by (\ref{eq:modelmatC}) for instance).

\section{Sensitivity analysis and risk allocation}
\label{sec:ana}%
Risk allocation consists in decomposing a portfolio risk measure (here, the expected or unexpected loss) into a sum of risk contributions by sub-portfolios (the sub-portfolios can be the groups discussed above in the paper or other groups, provided the portfolio loss can be expressed as (\ref{eq:portsub})).  
It makes it possible to determine the risk profile of the portfolio and to identify the most risky components of the portfolio.
Following \cite{denault} we use the Euler principle to obtain the decomposition.

The random loss of the portfolio is 
\begin{equation}
\label{eq:portsub}
L ({\bf Z})= \sum_{p=1}^P K_p \ell_p({\bf Z})  ,
\end{equation}
where $\ell_p({\bf Z}) $ is the random loss of sub-portfolio $p$ per unit principal 
and $K_p$ is the principal of sub-portfolio $p$.

The expected loss is
\begin{equation}
L^{\rm e} = \sum_{p=1}^P K_p \ell_p^{\rm e}  ,
\end{equation}
with $\ell_p^{\rm e}=\EE[  \ell_p({\bf Z})]$. 
We can then define the risk contribution, resp. sensitivity, of the expected loss to sub-portfolio $p$ by
\begin{equation}
RC_p^{\rm e} = K_p \ell_p^{\rm e},\quad \quad 
S_p^{\rm e} = \frac{K_p \ell_p^{\rm e}}{\sum_{p'=1}^P K_{p'} \ell_{p'}^{\rm e}}  .
\end{equation}

The unexpected loss is
\begin{equation}
L^{\rm u} \mbox{ such that } \PP( L ({\bf Z})\leq L^{\rm u})=1-\alpha  .
\end{equation}
Applying Euler's theorem to the homogeneous function of degree one $(K_p)_{p=1}^P \mapsto L^{\rm u}$ (see Appendix \ref{app:euler}),  
we obtain
\begin{equation}
L^{\rm u} = \sum_{p=1}^P K_p \partial_{K_p} L^{\rm u}   ,
\end{equation}
and (see Appendix \ref{app:euler}) we have $\partial_{K_p} L^{\rm u} =\EE[ \ell_p({\bf Z}) | L({\bf Z})=L^{\rm u}]$.   We can then define the risk contribution, resp. sensitivity, of the unexpected loss to sub-portfolio $p$ by
\begin{equation}
RC_p^{\rm u} = K_p \EE[ \ell_p({\bf Z}) | L({\bf Z})=L^{\rm u}] ,\quad \quad 
S_p^{\rm u} = \frac{K_p \EE[ \ell_p({\bf Z}) | L({\bf Z})=L^{\rm u}]}{\sum_{p'=1}^P K_{p'} \EE[ \ell_{p'} ({\bf Z}) | L({\bf Z})=L^{\rm u}]}  .
\end{equation}

These  risk contributions  and sensitivity indices can be estimated as follows.
Let us assume that we have a Monte Carlo sample $(\ell_p({\bf Z}^{(k)}))_{p=1}^P$ of size $N_{\rm MC}$ that 
is independent and identically distributed as $(\ell_p({\bf Z}))_{p=1}^P$.\\
We can estimate $L^{\rm e}$ by the empirical mean
$$
\hat{L}^{\rm e} = \frac{1}{N_{\rm MC}} \sum_{k=1}^{N_{\rm MC}}  \sum_{p=1}^P K_p \ell_p({\bf Z}^{(k)}),
$$
and we can estimate $RC_p^{\rm e}$ and $S^{\rm e}_p$ by
$$
\widehat{RC}^{\rm e}_p = 
\frac{1}{N_{\rm MC}} \sum_{k=1}^{N_{\rm MC}} K_p \ell_p({\bf Z}^{(k)}) ,\quad \quad 
\hat{S}^{\rm e}_p = 
\frac{ \sum_{k=1}^{N_{\rm MC}} K_p \ell_p({\bf Z}^{(k)})
}{
\sum_{k=1}^{N_{\rm MC}} \sum_{p'=1}^P K_{p'} \ell_{p'}({\bf Z}^{(k)})
}  .
$$

We can estimate $L^{\rm u}$ by the $1-\alpha$-empirical quantile $\hat{L}^{\rm u}$ of the sample  $(L^{(k)})_{k=1}^{N_{\rm MC}}$, with $L^{(k)}=\sum_{p=1}^P K_p \ell_p({\bf Z}^{(k)}) $.
We can estimate $RC^{\rm u}_p$ and $S^{\rm u}_p$ by
\begin{equation}
\widehat{RC}^{\rm u}_p = 
K_p \hat{s}_p,\quad \quad
\hat{S}^{\rm u}_p = \frac{ 
K_p \hat{s}_p
}{
\sum_{p'=1}^P K_{p'} \hat{s}_{p'}  
}  ,
\end{equation}
where $\hat{s}_p$ is an estimator of $\EE[ \ell_p({\bf Z}) | L({\bf Z})=L^{\rm u}]$.
We can use the Nadaraya-Watson estimator \cite[Chapter 6]{hastie}:
$$
\hat{s}_p
 =
 \frac{
 \sum_{k=1}^{N_{\rm MC}} \ell_p({\bf Z}^{(k)}) {\cal K}_h(   L^{(k)} - \hat{L}^{\rm u} )
 }
 {
 \sum_{k=1}^{N_{\rm MC}}  {\cal K}_h(   L^{(k)} - \hat{L}^{\rm u} )
 }  ,
 $$
 where ${\cal K}_h(x) =  {\cal K}(\frac{x}{h})$, with ${\cal K}(x) = \frac{1}{\sqrt{2\pi}}\exp(-\frac{x^2}{2})$ the Gaussian kernel (we could take the Epanechnikov quadratic kernel) and $h$ the bandwidth.

\section{Perspectives}
\label{sec:rev}%

\subsection{Reverse stress test}
The strategy developed in Section \ref{sec:condloss}
makes it possible to estimate the conditional loss under stressed conditions $L_{\rm stress}^{1-\alpha}$.
We may want to determine which systematic risk is the most important or which types of trajectories are the most likely to lead to 
a loss that exceeds $L_{\rm stress}^{1-\alpha}$.
For this, we can look for the conditional distribution of the process ${\bf Z}$ given $L({\bf Z})\geq L_{\rm stress}^{1-\alpha}$.
We may in particular want to determine $\EE [ \bZ_t | L({\bf Z})\geq L_{\rm stress}^{1-\alpha}]$ for $t=1,\ldots,T$.
This could be estimated by a straightforward use of the Monte Carlo sample generated for the estimation of the quantile $L_{\rm stress}^{1-\alpha}$.

\subsection{Systematic risk factor models}
In this paper we have proposed a climate-extended portfolio credit risk model that gives portfolio loss distributions conditional to a climate scenario.
If one wants to probabilize the scenarios, then one would need a model for the joint process of the systematic risk factors.
This could be possible within the framework proposed in this paper if the process is Gaussian.
In other words, it could be possible to design a parametric VAR(1) model  (vector auto-regressive) for ${\bf Z}$ and to exploit it with the general credit risk modeling developed in the paper. This work (including the calibration aspects) is under progress \cite{jules}.

\subsection{Climatic idiosyncratic risk models}
In this paper we have modeled the transition risk and physical risk as systematic risk factors. 
This seems well motivated for the transition risk that affects globally all borrowers.
We may think, however, that the physical risks may have systematic and idiosyncratic components. In such a case, the macro-correlations $(\zeta_{t,2+r})_{r=1}^R$ given by the climate scenario could also affect the variances of the idiosyncratic risks that are assumed to be time-independent in the paper. It is possible to incorporate such an effect within the framework proposed in this paper because the Gaussian copula model would not be affected.

\subsection{Granularity adjustment}
If the non-concentration condition (\ref{eq:ncc}) is not fulfilled,
then granularity adjustment is necessary 
to take into account that the portfolio may carry some undiversified idiosyncratic risk.
Granularity adjustment can be carried out by a Monte Carlo approach or by an analytical approach (through Taylor series expansions of the quantiles) and it has been the subject of intense research for the single-factor model \cite{wilde,vasicek,pykthin02,gordy03,emmer,gordy13,roncalli} and also for multi-factor models \cite{gordy03,egloff,fermanian}.
It would require some more work to get an appropriate version for the multi-factor model addressed in this paper.\\

%\clearpage
\section{Application}
\label{sec:app}%
In this section we apply the CERM-based method to estimate the portfolio loss distributions of a pilot portfolio for three climate scenarios.

\begin{figure}
\centering
\includegraphics[width=1.1\textwidth]{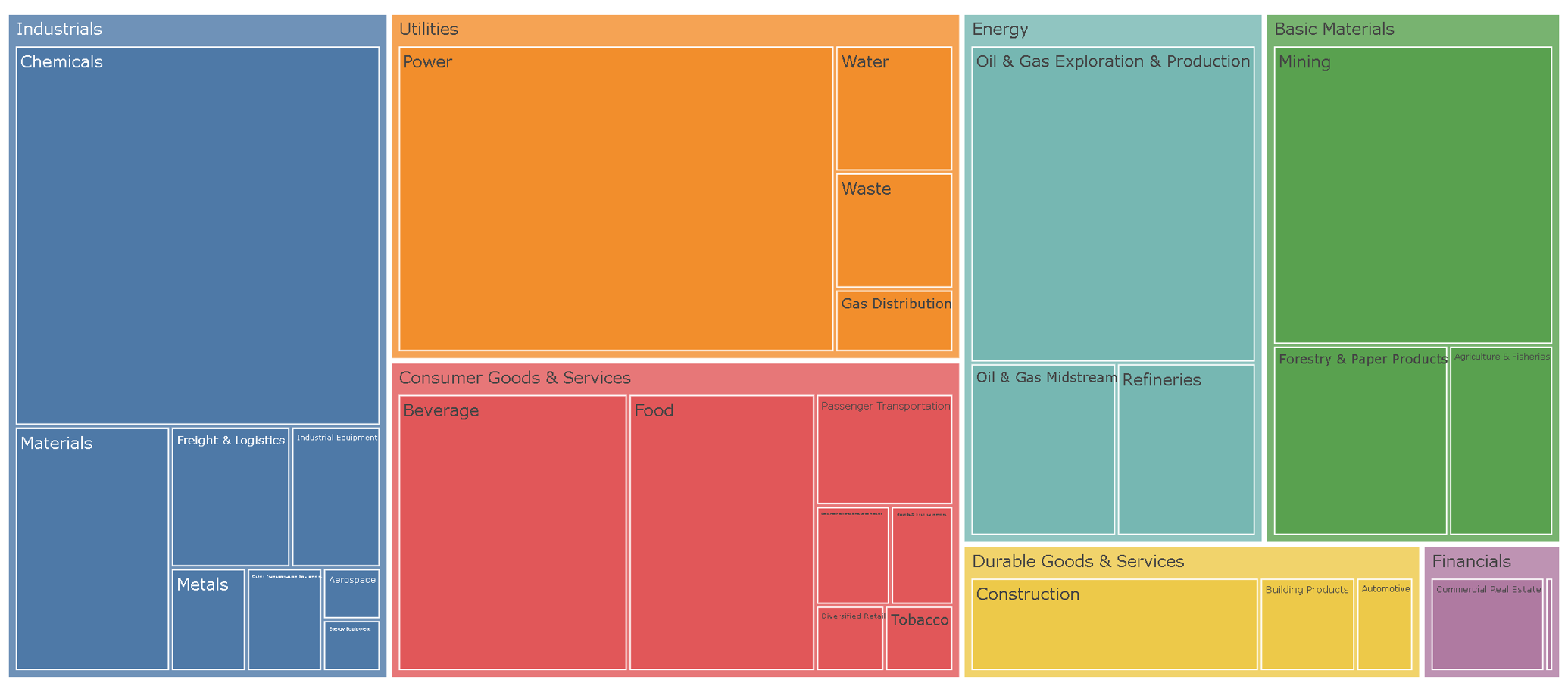}
\caption{Pilot portfolio, heatmap with the EAD per sector and subsector (the EAD is proportional to the area of the sector in this figure).}
\label{fig:app1}
\end{figure}

\begin{figure}
\centering
\includegraphics[width=1.1\textwidth]{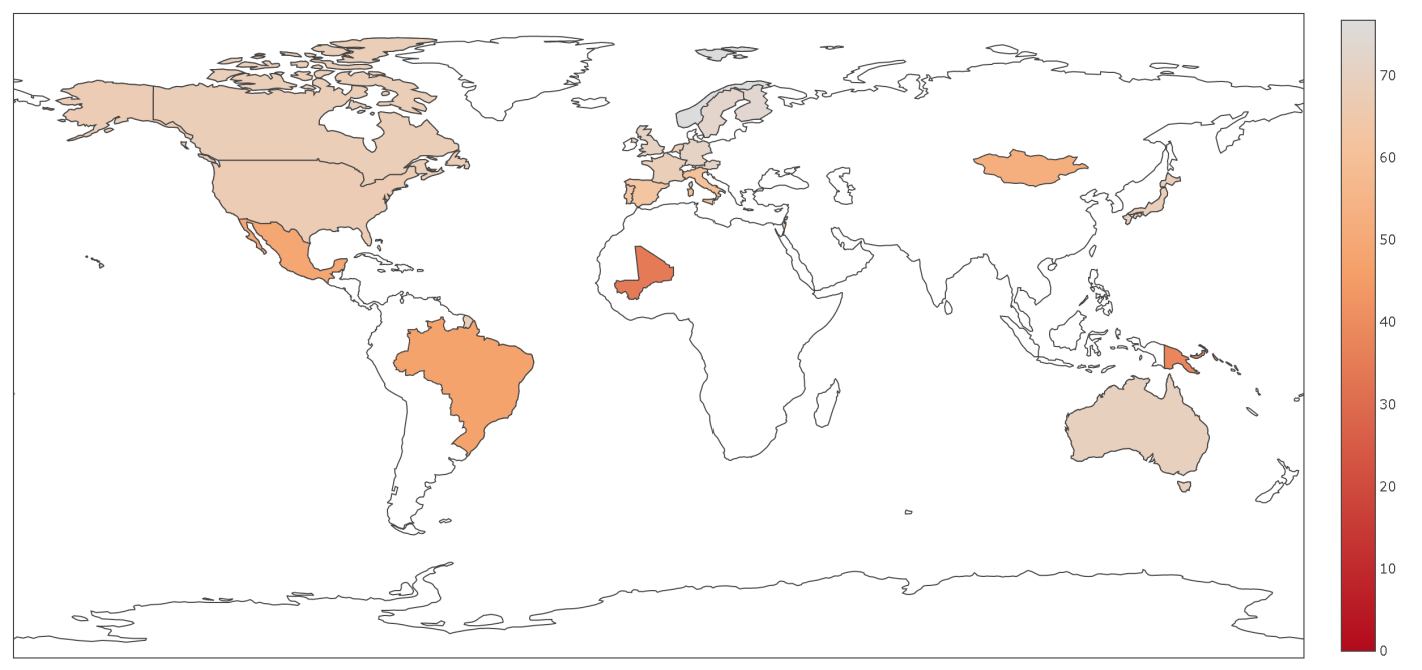}
\caption{Pilot portfolio, geographic distribution of the EAD. The colour coding depends upon the vulnerability to climate change of the borrower's main country using ND Gain data sets (see https://gain.nd.edu/our-work/country-index/).}
\label{fig:app2}
\end{figure}

%\subsection{Pilot portfolio}
The book is made of corporate loans of borrowers from various sectors and sub-sectors spread across regions (see Figs.~\ref{fig:app1}-\ref{fig:app2}). 

\begin{figure}
\includegraphics[width=\textwidth]{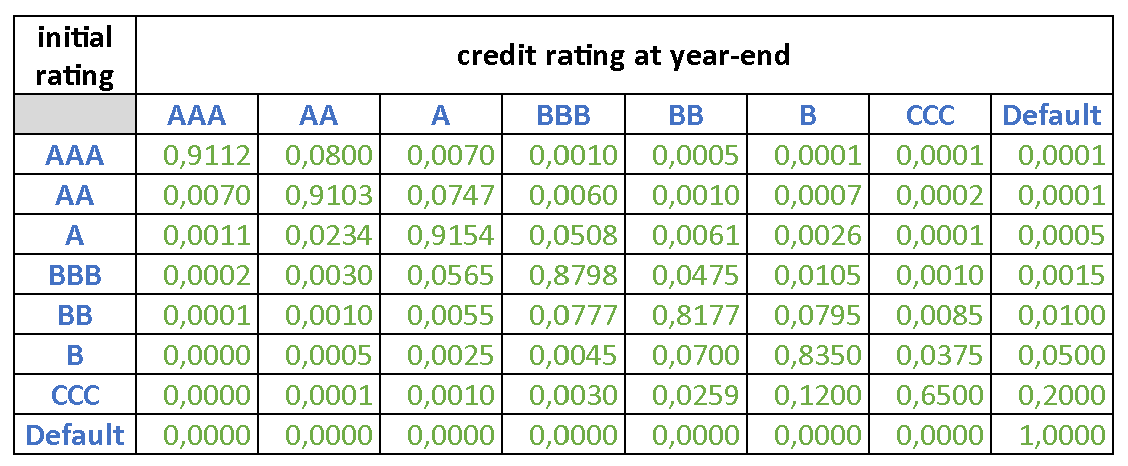}
\caption{$1$-year migration matrix with $K=8$. Each row corresponds to an initial rating. Each column corresponds to a rating at the end of one year. As "Default" is absorbing, the last line is of the form $(0,\ldots,0,1)$.}
\label{fig:matrat}
\end{figure}

%\subsection{Credit data}
We use credit ratings from Standard \& Poor's on a simplified scale with $K=8$ and the unconditional migration matrix given in Fig.~\ref{fig:matrat}. 

\begin{figure}
\centering
\includegraphics[width=1.1\textwidth]{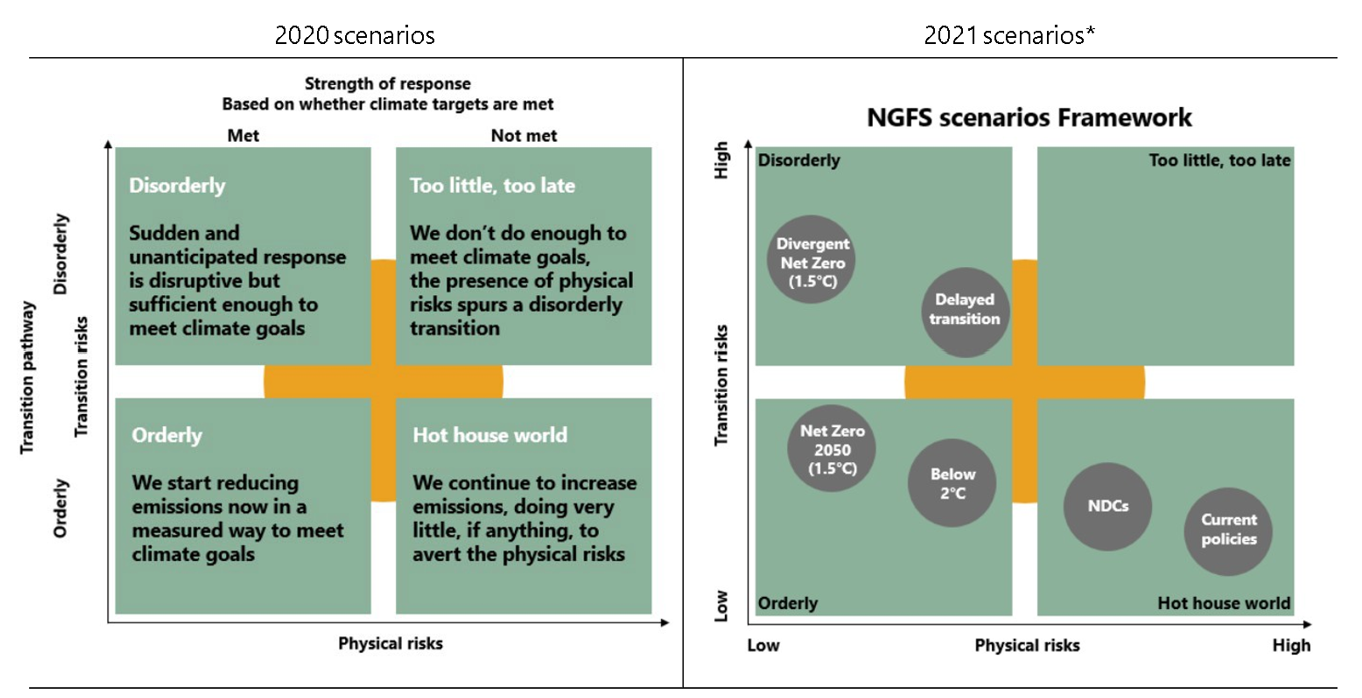}
\caption{NGFS scenarios. On the horizontal axis it is organized by climate outcomes. On the left we meet the climate targets set out in the Paris agreement. On the right we fail to do so and get substantial physical risk. On the vertical axis we look at the type of transition. Orderly on the bottom row or disorderly on the top row, reflecting whether policy actions are taken early or late and whether technological progress is able to mitigate some of the costs. There are two successive versions 2020 and 2021.}
\label{fig:app3}
\end{figure}

\begin{figure}
\centering
\includegraphics[width=1.1\textwidth]{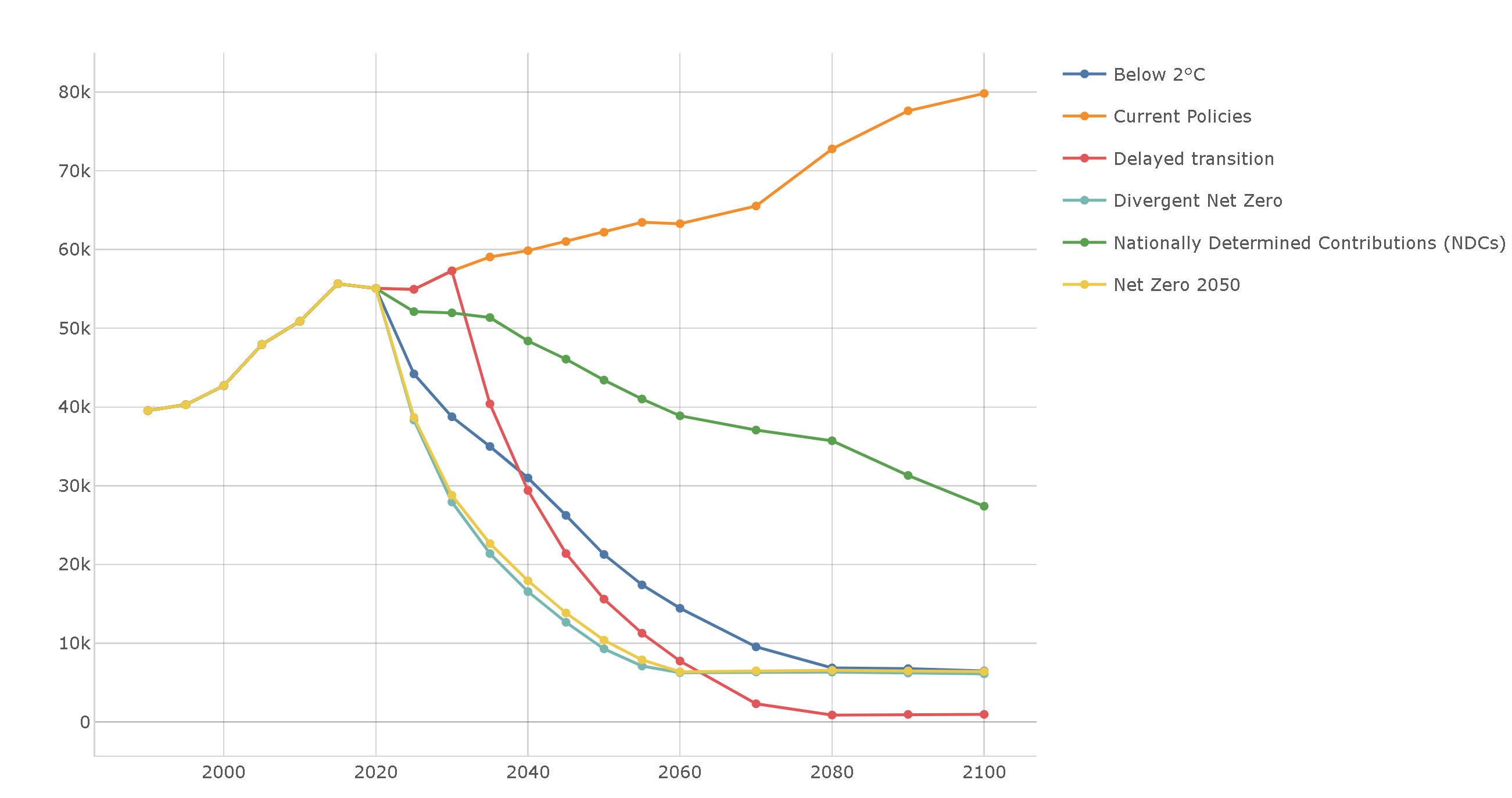}
\caption{NGFS GHG pathways expressed in CO2 equivalent emissions, in billion tons of CO2 per year.}
\label{fig:app3b}
\end{figure}

%\clearpage
%\subsection{Climate scenarios}
In this work we look at different possible outcomes, looking forward up to the end of the century. We take climate scenarios from the NGFS framework. The NGFS is the Network for Greening the Financial System, a global network of 100 central banks and supervisors. The NGFS developed a matrix to classify scenarios (see Fig.~\ref{fig:app3}). We consider three climate scenarios from the framework : Orderly (Below 2°C), Disorderly (Delayed transition) and Hot House World (Current policies). Each scenario is characterized by a temperature rise, a GHG budget and a pathway to spend this budget. For instance, to reach a temperature increase of less than 2°C by 2100, we have to go from circa 55 gigatons of CO2 equivalent per year to 0 within the next 30 years. This implies massive transformations across all areas of the economy leading to massive value adjustments. This risk has very material impacts on the 2020-2050 time window (see Fig.~\ref{fig:app3b}).

\begin{figure}[t]
\centering
    \begin{subfigure}[b]{0.8\textwidth}
        \includegraphics[width=\textwidth]{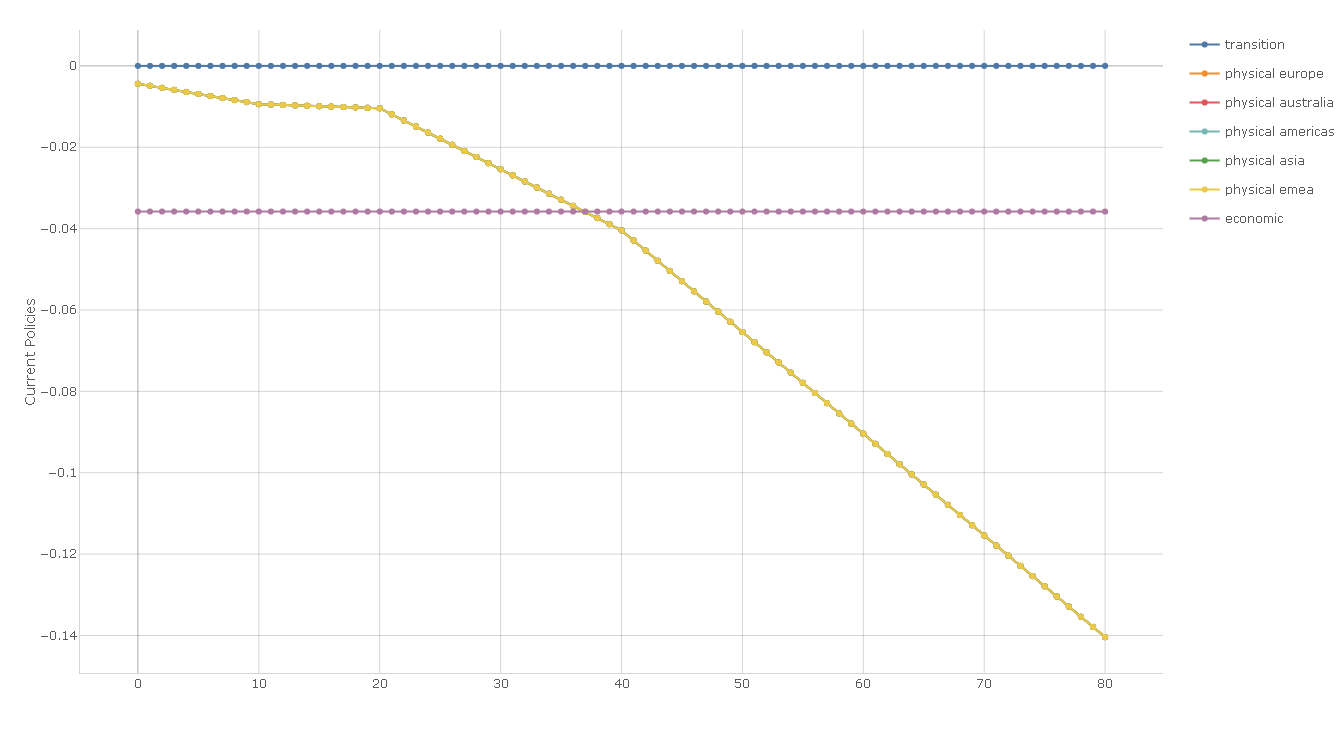}
    \end{subfigure}
    \begin{subfigure}[b]{0.8\textwidth}
        \includegraphics[width=\textwidth]{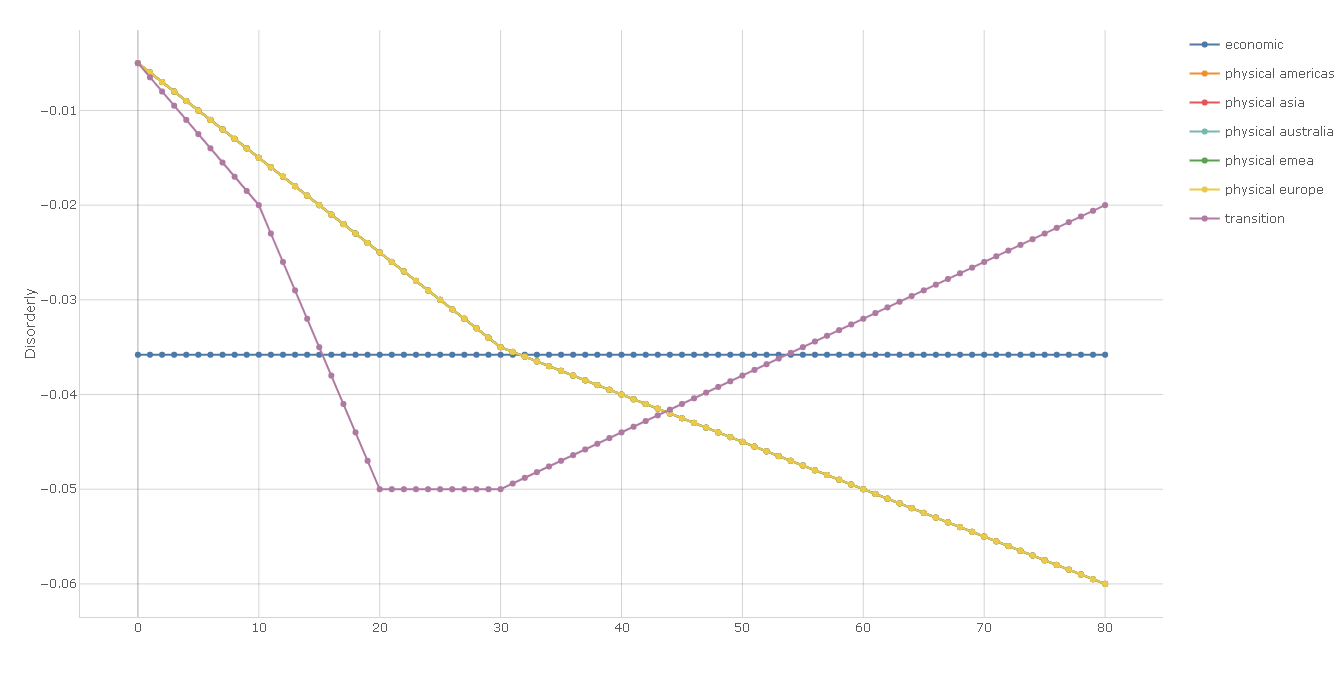}
    \end{subfigure}
    \begin{subfigure}[b]{0.8\textwidth}
        \includegraphics[width=\textwidth]{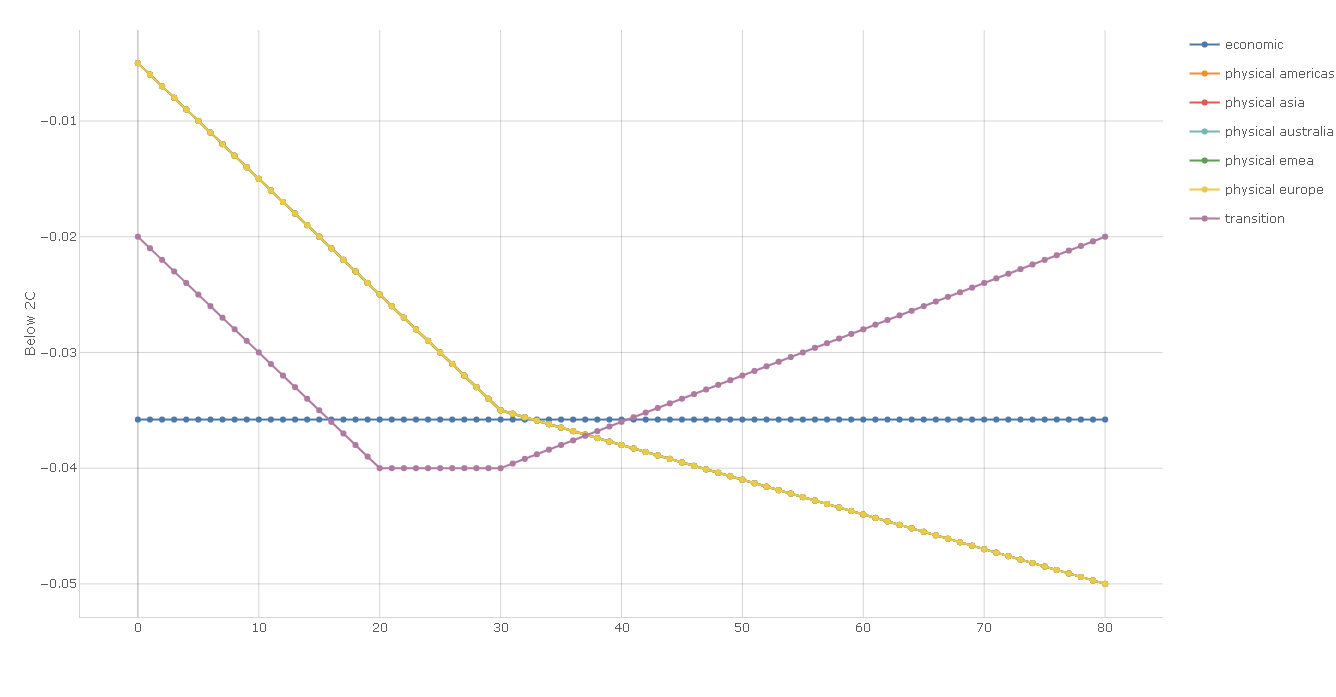}
    \end{subfigure}
\caption{Example of macro-correlation parameters from 2020 to 2100 for three NGFS climate scenarios: Below 2°C (top), Disorderly (center), and Current Policies (bottom). 
The blue lines stand for the economic risk intensity,
the magenta lines stand for the transition risk intensity, and
the yellow lines stand for the physical risks intensity (all physical risks share the same intensity).}
\label{fig:app4}
\end{figure}

We apply the CERM-based method with the macro-correlations expressed in GDP growth rate determined by these scenarios (see Fig.~\ref{fig:app4}).
In the three scenarios the economic risk intensity is constant. In the Orderly scenario the transition starts now in a progressive way. In the Delayed scenario the transition occurs later with a more disruptive impact. In the Hot House World scenario the transition risk is zero as there is almost no regulation, but the physical risk intensity dramatically increases.

\begin{figure}
\centering
\includegraphics[width=\textwidth]{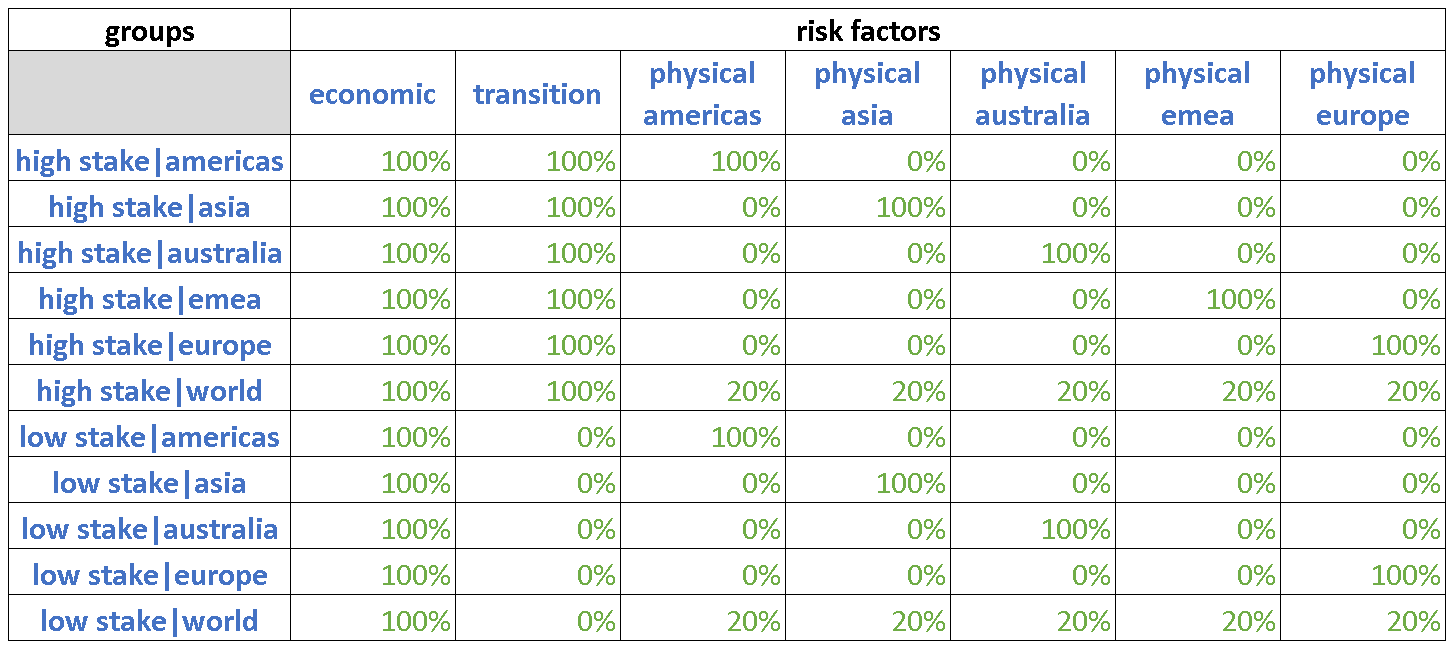}
\caption{Micro-correlation parameters by group. In this simplified version the most difficult borrowers when it comes to low-carbon transition are classified as "high stake", the others are classified as "low stake". Physical risk impact is linked to the borrower's main country, either in terms of location or of market, depending on the sector. Sector specificities and sub-country spatial resolution are introduced in more advanced setup.}
\label{fig:app5}
\end{figure}

We use flat micro-correlations across time using the simplified grid with $13$ groups given in Fig.~\ref{fig:app5}. For a more granular analysis micro-correlation parameters can be calibrated to GHG intensity data sets from specialised providers such as Carbon4 Finance\footnote{https://www.carbon4finance.com/} or MSCI\footnote{https://www.msci.com/}. Micro-correlation parameters can also evolve with time to account for mitigation and adaptation Capex plans. 

\begin{figure}
\centering
\includegraphics[width=0.6\textwidth]{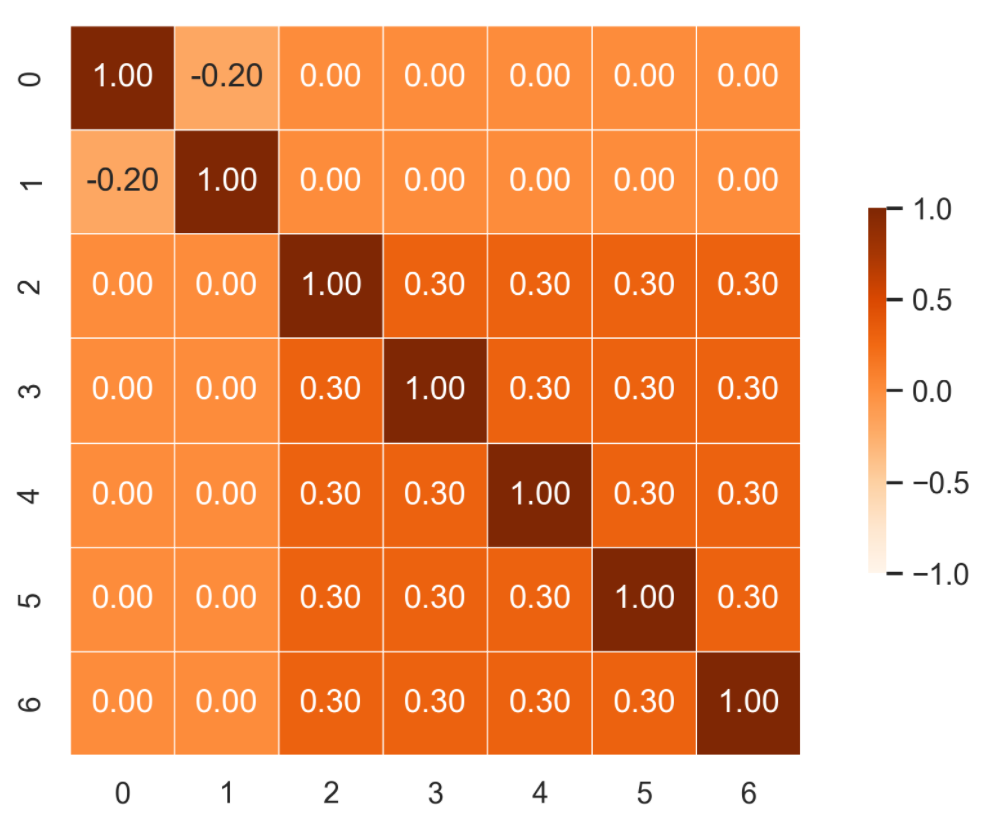}
\caption{Covariance matrix for the $7$ risk factors with an anti-correlation between the economic risk and the transition risk and a positive correlation between the 5 regional risk factors (here the index $0$ corresponds to the economic risk, $1$ corresponds to the transition, and $2$--$6$ correspond to the $5$ regions: Europe, America, Asia, Australia, Middle East \& Africa).}
\label{fig:app6}
\end{figure}

We use $7$ risk factors: economic risk, transition risk, and physical risk for $5$ regions. The physical risk factors have the same intensity (plotted in Fig.~\ref{fig:app4}). Their covariance matrix is given in Fig.~\ref{fig:app6}.

%\clearpage

We can easily convert the expected loss into a credit spread impact. We do the same for the cost to maintain buffers against unexpected loss quantiles. We use a parameter which is the cost to raise sufficient capital from shareholders, thus providing extra loss absorption capacity, 10pct in this example. We get an overall impact from the climate on credit spreads, which we call the climate risk premium. 

\begin{figure}
\centering
    \begin{subfigure}[b]{0.49\textwidth}
        \includegraphics[width=\textwidth]{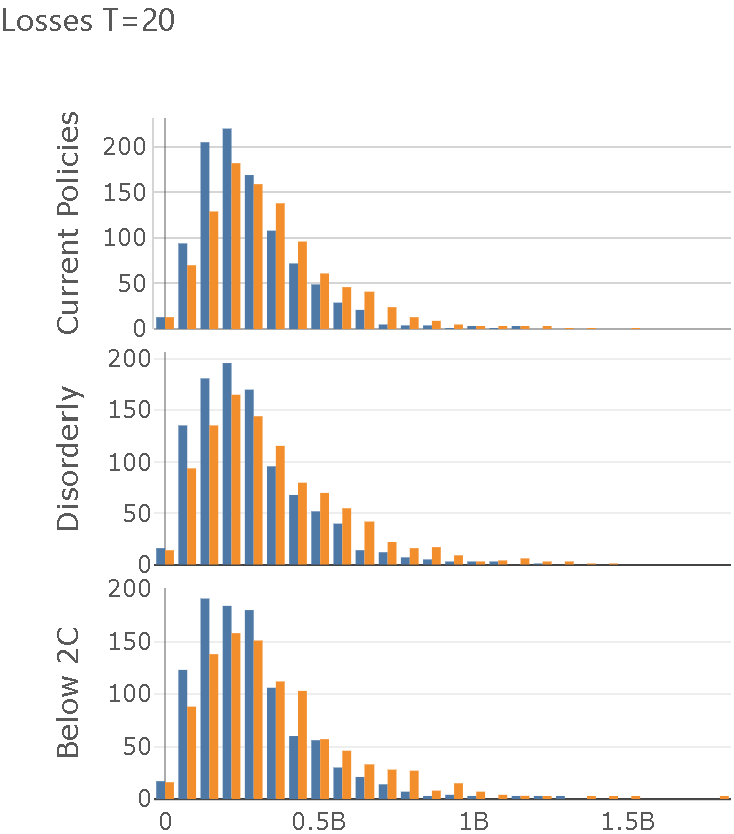}
    \end{subfigure}
    \begin{subfigure}[b]{0.49\textwidth}
        \includegraphics[width=\textwidth]{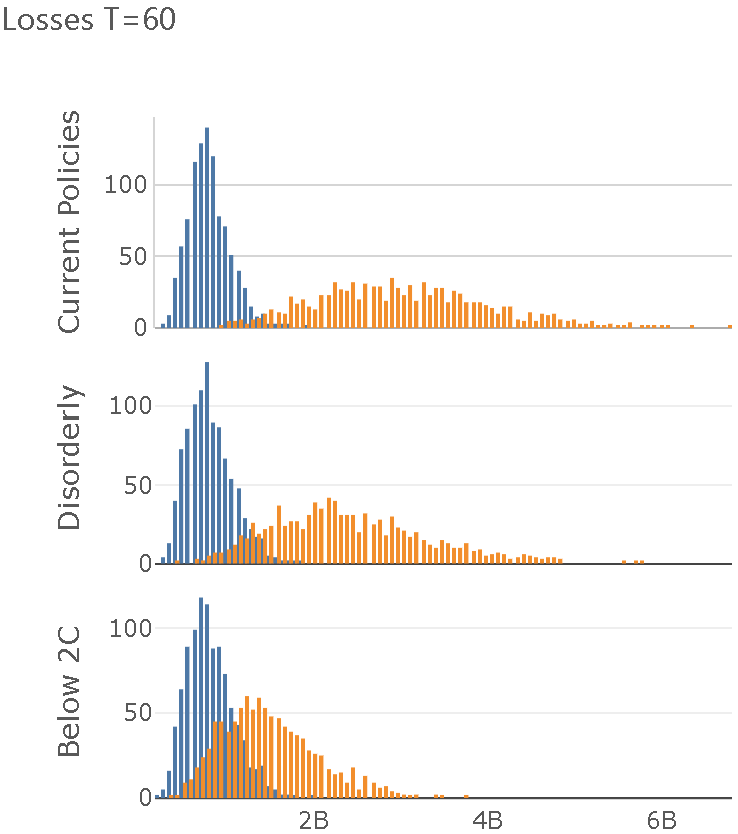}
    \end{subfigure}
\caption{Loss distributions with the CERM-based approach on the pilot portfolio and on the three NGFS climate scenarios: Current Policy (top), Disorderly (center), and Below 2°C (bottom). In the left column the horizon is $2042$, in the right column it is $2082$. The blue distributions are obtained with only economic risk, in absence of transition and physical risks, and are, therefore, identical for the three scenarios (up to Monte Carlo fluctuations).
The orange distributions are obtained in presence of economic, transition and physical risks.}
\label{fig:app7}
\end{figure}

\begin{figure}
\centering
\includegraphics[width=1\textwidth]{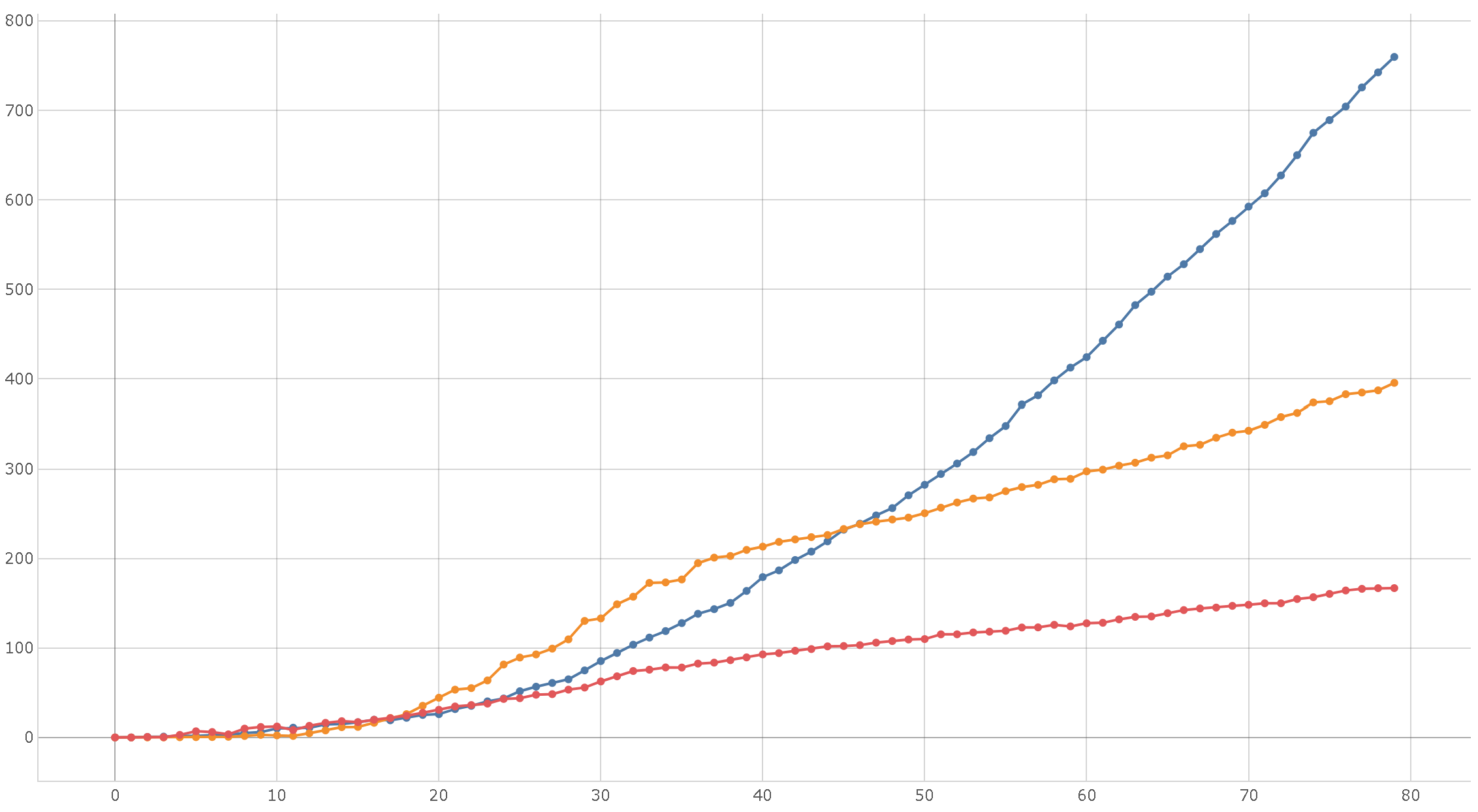}
\caption{Climate risk premium in Basis Points per time horizon on the same sample portfolio, computed as the sum of $L^{\rm e}$ and the cost of additional capital from shareholders times $L_{\rm stress}^{1-\alpha}$ and on the three NGFS climate scenarios: Current Policy (blue), Disorderly (orange), and Below 2°C (red)}
\label{fig:app8}
\end{figure}

\begin{figure}
\centering
\includegraphics[width=\textwidth]{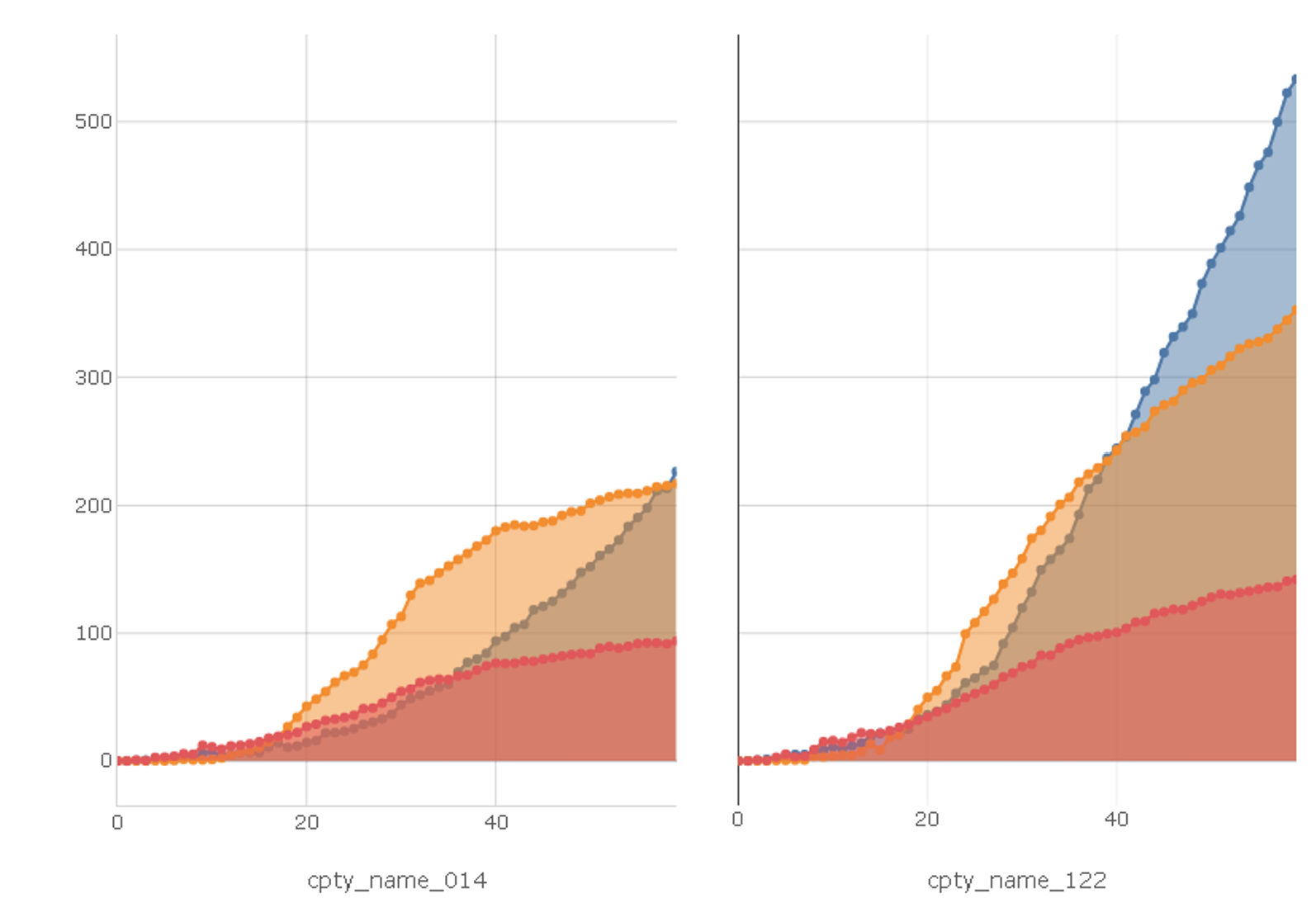}
\caption{Climate risk premium in Basis Points for two obligors on the three NGFS climate scenarios: Current Policy (blue), Disorderly (orange), and Below 2°C (red). The horizon is 60 years. The first obligor belongs to the "low stake" category, the second one belongs to the 'high stake" category.}
\label{fig:app9}
\end{figure}

The portfolio loss distributions and the climate risk premiums for one particular sector are plotted in Figs.~\ref{fig:app7} and \ref{fig:app8} for the three climate scenarios of Fig.~\ref{fig:app4}. As can be easily understood, if the Hot House World scenario seems to be reasonable in the short term (2050), it becomes unsustainable in the long term (2100). 
In Fig.~\ref{fig:app9} we can compare side by side obligors and see the impact from different micro-correlations. 

\clearpage

\section*{Acknowledgements}
This work was initiated by the association Green RWA (Risk Weighted Assets). It was written in collaboration with Olivier Vinciguerra (cerm@greenrwa.org), who leads the Association. It extends the model proposed in a first white paper published by Green RWA \cite{rwa}. Iggaak took care of implementing the model code and provided illustrative metrics on a pilot portfolio.
The interested readers are invited to contact the authors and/or the association for questions, comments, and suggestions.

We thank Emmanuel Gobet and the members of the Chair Stress Test, RISK Management and Financial Steering (a research program  between Ecole Polytechnique and BNP Paribas) for useful and stimulating discussions.

\appendix

\section{The special recovery model (\ref{eq:lossrec})}

In this appendix we give more detailed results about the recovery model (\ref{eq:lossrec}) which is inspired from \cite{andersen}. This model allows for flexibility, easy manipulation, and (relatively) easy calibration.
It uses the cumulative Gaussian distribution function $\Phi$ for easy calculations, by the two following Gaussian formulas:
\begin{align}
\frac{1}{\sqrt{2\pi}} 
\int_{-\infty}^\infty \Phi(ax+b) \exp\big(- \frac{x^2}{2}\big) dx &= \Phi\Big( \frac{b}{\sqrt{1+a^2}}\Big)  ,\\
\frac{1}{\sqrt{2\pi}} 
\int_{-\infty}^c \Phi(ax+b) \exp\big(- \frac{x^2}{2}\big) dx &= \Phi_2\Big(\frac{b}{\sqrt{1+a^2}} ,c; - \frac{a}{\sqrt{1+a^2}}\Big) ,
\end{align}
where $\Phi$ is the cdf of the standard Gaussian distribution and $\Phi_2(\cdot,\cdot;\rho)$ is the bivariate cumulative Gaussian distribution with correlation $\rho$.

Eq.~(\ref{eq:lgdhist0}) (and also (\ref{eq:tauXR0})) can be used to calibrate the parameters of the recovery model from default swap market data. 
More elaborate moment matching can be used because it is also possible to express, in simple closed forms, 
the moments $\EE\big[ \Phi^{-1}({\rm RR}_{t}^{(q)})^n | X_{t}^{(q)} \leq z_{{g},t,iK} \big]$
of the recovery rate for a borrower with rating $i$ at time $t-1$ who defaults at time $t$ (see Appendix \ref{app:12}).

\subsection{The rank correlation}
\label{app:11}
The Kendall's Tau (rank correlation) of a pair of random variables $(X,Y)$ is defined by
\begin{equation}
\tau(X,Y) = \PP\big( (X-\tilde{X})(Y-\tilde{Y}) >0 \big)- \PP\big( (X-\tilde{X})(Y-\tilde{Y}) < 0 \big),
\end{equation}
when $(X,Y)$ and $(\tilde{X},\tilde{Y})$ are independent and identically distributed.

From (\ref{eq:asset}) and (\ref{eq:loss})-(\ref{eq:lossrec}), we get that the Pearson (linear) correlation coefficient between $X$ and $\Phi^{-1}(R)$ is 
\begin{equation}
\rho(X_{t}^{(q)},\Phi^{-1}({\rm RR}_{t}^{(q)})) = 
 \ba_{{g},i,t}\cdot {\bf C} \bb_{{g},i,t}   .
\end{equation}
Since $(X,\Phi^{-1}(R))$ is a Gaussian vector, the Kendall's Tau is related to the Pearson correlation coefficient through the Greiner's equality:
\begin{equation}
\tau(X_{t}^{(q)},\Phi^{-1}({\rm RR}_{t}^{(q)})) =  \frac{2}{\pi} {\rm arcsin}\big( \rho(X_{t}^{(q)},\Phi^{-1}({\rm RR}_{t}^{(q)})) \big).
\end{equation}
The Kendall's Tau is invariant under strictly increasing transform, so 
\begin{align}
\nonumber
\tau(X_{t}^{(q)},{\rm RR}_{t}^{(q)}) &=  
\tau(X_{t}^{(q)},\Phi^{-1}({\rm RR}_{t}^{(q)})) 
\\
& = 
\frac{2}{\pi} {\rm arcsin}\big( \ba_{{g},i,t}\cdot {\bf C} \bb_{{g},i,t}  \big).
\label{eq:tauXR}
\end{align}

\subsection{The Loss Given Default}
\label{app:12}
 It is important to note (for calibration purposes) that the recovery rate and the default occurrence are correlated.
This means that the unconditional expectation of (one minus) the recovery rate 
\begin{equation}
\EE  \big[ 1- {\rm RR}_{t}^{(q)} \big] 
=
1-\Phi \Big( \frac{\mu_{{g},i,t}   }{ \sqrt{1+\sigma_{{g},i,t}^2}}\Big)   
\end{equation}
is not the expected Loss Given Default, that is observed for the borrowers who default.
The expected Loss Given Default for the borrowers from group ${g}$ and with rating $i$ at time $t-1$ who default at time $t$ is
\begin{equation}
\EE  \big[ 1- {\rm RR}_{t}^{(q)} | X_{t}^{(q)} \leq z_{{g},t,iK} \big]  ,
\end{equation}
because the event ``$X_{t}^{(q)} \leq z_{{g},t,iK}$" corresponds to default for such borrowers.
The  expected Loss Given Default  for the borrowers from group ${g}$ and with rating $i$ at time $t-1$ 
who default at time $t$ actually depends on the rating $i$:
\begin{align}
\label{eq:lgdhist}
\EE  \big[ 1- {\rm RR}_{t}^{(q)} | X_{t}^{(q)} \leq z_{{g},t,iK} \big]
&  = 1-
\frac{1}{({\bf M}_{{g},t})_{iK}}
\Phi_2\Big( \frac{\mu_{{g},i,t}}{\sqrt{1+\sigma_{{g},i,t}^2 }} ,
z_{{g},t,iK} ; \frac{-\rho_{{g},i,t} \sigma_{{g},i,t} }{ \sqrt{1+\sigma_{{g},i,t}^2}}\Big)  ,
\end{align}
where $\rho_{{g},i,t}=  \ba_{{g},i,t}\cdot {\bf C} \bb_{{g},i,t} $.

{\it Proof.}
The distribution of $X_{t}^{(q)}$ is ${\cal N}(0,1)$.
The distribution of $\Phi^{-1}({\rm RR}_{t}^{(q)})$ is ${\cal N}(\mu_{{g},i,t}, \sigma_{{g},i,t}^2)$.
The correlation coefficient between $X_{t}^{(q)}$ and $\Phi^{-1}({\rm RR}_{t}^{(q)})$ is $\rho_{{g},t}$.
The vector $(X_{t}^{(q)},\Phi^{-1}({\rm RR}_{t}^{(q)}))$ is Gaussian, so the conditional distribution of $\Phi^{-1}({\rm RR}_{t}^{(q)})$
given $X_{t}^{(q)}=x$ is ${\cal N}(\mu_{{g},i,t}+\rho_{{g},i,t} \sigma_{{g},i,t} x, \sigma_{{g},i,t}^2 (1-\rho_{{g},i,t}^2))$ and we get 
\begin{align*}
\EE[{\rm RR}_{t}^{(q)}|X_{t}^{(q)}=x] &= 
 \frac{1}{\sqrt{2\pi \sigma_{{g},i,t}^2 (1-\rho_{{g},i,t}^2)}}
\int_{-\infty}^\infty \Phi(r)
\exp\Big( - \frac{(r- \mu_{{g},i,t}- \rho_{{g},i,t} \sigma_{{g},i,t} x)^2}{2 \sigma_{{g},i,t}^2 (1-\rho_{{g},i,t}^2)}\Big) dr 
\\
&=
\Phi\Big( \frac{\mu_{{g},i,t}+\rho_{{g},i,t}\sigma_{{g},i,t} x}{\sqrt{1+\sigma_{{g},i,t}^2(1-\rho_{{g},i,t}^2)}}\Big)
\end{align*}
and
\begin{align*}
\EE  \big[ {\rm RR}_{t}^{(q)} | X_{t}^{(q)} \leq z_{{g},t,iK} \big]
&=
\frac{
\EE  \big[ {\rm RR}_{t}^{(q)} {\bf 1}_{ X_{t}^{(q)} \leq z_{{g},t,iK} } \big]}{\PP( X_{t}^{(q)} \leq z_{{g},t,iK})}\\
&=
\frac{
\frac{1}{\sqrt{2\pi}} 
\int_{-\infty}^{z_{{g},t,iK}}
\Phi\Big( \frac{\mu_{{g},i,t}+\rho_{{g},i,t}\sigma_{{g},i,t} x}{\sqrt{1+\sigma_{{g},i,t}^2(1-\rho_{{g},i,t}^2)}}\Big)
\exp\big(-\frac{x^2}{2}\big)
dx
}{({\bf M}_{{g},t})_{ik}}  ,
\end{align*}
which gives (\ref{eq:lgdhist}).
\qed

Similarly,
we have
\begin{align}
\label{eq:monPhiR1}
\EE  \big[ \Phi^{-1}({\rm RR}_{t}^{(q)}) | X_{t}^{(q)} \leq z_{{g},t,iK} \big]
=&  \mu_{{g},i,t} - \rho_{{g},i,t} \sigma_{{g},i,t} \frac{ \exp( - z_{{g},t,iK}^2/2)}{\sqrt{2\pi} ({\bf M}_{{g},t})_{iK}}
,
 \\
\nonumber
\EE  \big[ \Phi^{-1}({\rm RR}_{t}^{(q)})^2 | X_{t}^{(q)} \leq z_{{g},t,iK} \big]
=&\mu_{{g},i,t}^2-2 \rho_{{g},i,t} \sigma_{{g},i,t} \mu_{{g},i,t} 
\frac{ \exp( - z_{{g},t,iK}^2/2)}{\sqrt{2\pi} ({\bf M}_{{g},t})_{iK}}
\\
&-
\sigma_{{g},i,t}^2 \rho_{{g},i,t}^2
z_{{g},t,iK} \frac{ \exp( - z_{{g},t,iK}^2/2)}{\sqrt{2\pi} ({\bf M}_{{g},t})_{iK}}
+
\sigma_{{g},i,t}^2
.
\label{eq:monPhiR2}
\end{align}

{\it Proof.}
These formulas follow from the fact that the conditional distribution of $\Phi^{-1}({\rm RR}_{t}^{(q)})$
given $X_{t}^{(q)}=x$ is ${\cal N}(\mu_{{g},i,t}+\rho_{{g},i,t} \sigma_{{g},i,t} x, \sigma_{{g},i,t}^2 (1-\rho_{{g},i,t}^2))$,
so that
\begin{align*}
\EE  \big[ \Phi^{-1}({\rm RR}_{t}^{(q)}) | X_{t}^{(q)} =x \big]
=&\mu_{{g},i,t}+\rho_{{g},i,t} \sigma_{{g},i,t} x , \\
\EE  \big[ \Phi^{-1}({\rm RR}_{t}^{(q)})^2 | X_{t}^{(q)} \leq z_{{g},t,iK} \big]
=& \mu_{{g},i,t}^2+2 \rho_{{g},i,t} \sigma_{{g},i,t} \mu_{{g},i,t} x \\
&+ \rho_{{g},i,t}^2 \sigma_{{g},i,t}^2 x^2 +  \sigma_{{g},i,t}^2 (1-\rho_{{g},i,t}^2) .
\end{align*}
We get the desired results by using the following Gaussian identities:
\begin{align*}
\frac{1}{\sqrt{2\pi }} \int_{-\infty}^{z_{{g},t,iK}} \exp\big(-\frac{x^2}{2}\big) dx & = ({\bf M}_{{g},t})_{iK} ,\\
\frac{1}{\sqrt{2\pi }} \int_{-\infty}^{z_{{g},t,iK}} x \exp\big(-\frac{x^2}{2}\big) dx &= -  \frac{1}{\sqrt{2\pi }}
\exp \big( -\frac{z_{{g},t,iK}^2}{2}\big),  \\
\frac{1}{\sqrt{2\pi }} \int_{-\infty}^{z_{{g},t,iK}} x^2 \exp\big(-\frac{x^2}{2}\big) dx &= 
 -  \frac{z_{{g},t,iK}}{\sqrt{2\pi }}
\exp \big( -\frac{z_{{g},t,iK}^2}{2}\big) +
({\bf M}_{{g},t})_{iK} .
\end{align*}
\qed

The formulas (\ref{eq:tauXR}), (\ref{eq:lgdhist}), (\ref{eq:monPhiR1}), and (\ref{eq:monPhiR2}) 
can be used to calibrate the parameters $\mu_{{g},i,t}$, $\sigma_{{g},i,t}$, and $\bb_{{g},i,t}$ (or $\lambda_{{g},i,t}$ if we use the simplified model $\bb_{{g},i,t}=\lambda_{{g},i,t} \ba_{{g},i,t}$) of the recovery model.

The conditional Loss Given Default for the borrowers from group ${g}$ and with rating $i$ at time $t-1$  who default at time $t$ given $\bZ_t$ is simple, because the correlation between the recovery rate and the default occurrence happens only through the systematic risk factors,
so $X_{t}^{(q)}$ and ${\rm RR}_{t}^{(q)}$ are independent given $\bZ_t$. 
The conditional  loss given default for the borrowers from group ${g}$ and with rating $i$  who default at time $t$ given $\bZ_t$ is given by~(\ref{eq:lgdcond}):
\begin{align}
\nonumber
{\rm LGD}_{{g},i,t}(\bZ_t)  &:= \EE  \big[ 1- {\rm RR}_{t}^{(q)} | X_{t}^{(q)} \leq z_{{g},t,iK}, \bZ_t \big] \\
\nonumber
& =
\EE  \big[ 1- {\rm RR}_{t}^{(q)} | \bZ_t \big]\\
& =
 1 -\Phi \Big( \frac{\mu_{{g},i,t} +  \sigma_{{g},i,t} \bb_{{g},i,t}\cdot\bZ_t }{\sqrt{1+\sigma^2_{{g},i,t} (1-\bb_{{g},i,t}\cdot {\bf C} \bb_{{g},i,t})}}\Big)  .
\end{align}

 \section{Euler allocation principle}
 \label{app:euler}
Let $\bX=(X_1,\ldots,X_P)$ be a random vector, $\bw=(w_1,\ldots,w_P)$ a deterministic vector.
We introduce the random variable $L=\bw \cdot \bX =\sum_{p=1}^P w_p X_p$. We denote by $L^\alpha(\bw)$ the $1-\alpha$-quantile of $L$. The following statements then hold true:\\
1) $\bw \mapsto L^\alpha(\bw)$ is a homogeneous function of degree one.\\
2) For any $p=1,\ldots,P$:
\begin{equation}
\frac{\partial L^\alpha}{\partial w_p} (\bw)= \EE \big[ X_p | L=L^\alpha(\bw)] .
\end{equation}

{\it Proof.}
The proof is standard and we give it here for completeness.
We have $\PP( \bw \cdot \bX \leq L^\alpha(\bw))=1-\alpha$ and therefore
$\PP( c\bw \cdot \bX \leq c L^\alpha(\bw))=1-\alpha$,
which shows $L^\alpha(c\bw))=c L^\alpha(\bw)$. This proves the first item.\\
Let us denote by $f(\cdot;\bw)$, resp. $F(\cdot;\bw)$, the probability density function, resp.  the cumulative distribution function, of $L = \bw\cdot\bX$:
$$
F (z;\bw) = \PP( L \leq z)= \PP( \bw \cdot\bX \leq z).
$$
We have for all $\bw$: $F(L^\alpha(\bw);\bw)=1-\alpha$.
Therefore we have for any $p$:
$$
\partial_z F( L^\alpha(\bw);\bw) \partial_{w_p} L^\alpha(\bw) 
+ (\partial_{w_p}  F )(L^\alpha(\bw);\bw) = 0 ,
$$
and (since $\partial_z F=f$ and denoting $x_+=\max(x,0)$, $x_-=- \min(x,0)$, and $f_{X_p,L}$ the probability density function of $(X_p,L)$ assumed to be continuous)
\begin{align*}
&\partial_{w_p} L^\alpha(\bw)  = 
- \frac{ (\partial_{w_p}  F)(L^\alpha(\bw);\bw)  }
{ f( L^\alpha(\bw);\bw)  } \\
&
= - \lim_{\delta \to 0}
\frac{ F(L^\alpha(\bw);\bw +\delta {\itbf e}_p )
-F(L^\alpha(\bw);\bw) }
{ f( L^\alpha(\bw);\bw)  \delta}\\
&
= - \lim_{\delta \to 0}
\frac{ \PP( \bw \cdot\bX +\delta X_p \leq L^\alpha(\bw)) -\PP(  \bw \cdot\bX \leq L^\alpha(\bw)) }
{ f( L^\alpha(\bw);\bw  )  \delta}\\
&
=- \lim_{\delta \to 0}
\frac{\PP( L^\alpha(\bw)  \leq L \leq L^\alpha(\bw)  +\delta (X_p)_- )  - \PP( L^\alpha(\bw) - \delta (X_p)_+ \leq L \leq L^\alpha(\bw) )}
{ f( L^\alpha(\bw);\bw )  \delta}
\\
&
=- \lim_{\delta \to 0}
\int_{-\infty}^0 dx \int_{ L^\alpha(\bw) }^{ L^\alpha(\bw)  - \delta x} dz \frac{f_{X_p,L}(x,z;\bw )}{f(L^\alpha(\bw);\bw ) \delta}-
\int_0^{+\infty} dx \int_{ L^\alpha(\bw) -\delta x}^{ L^\alpha(\bw) } dz  \frac{f_{X_p,L}(x,z;\bw )}{f(L^\alpha(\bw);\bw ) \delta} \\
&
=
\int_{-\infty}^0 dx \, x \frac{f_{X_p,L}(x,L^\alpha(\bw);\bw )}{f(L^\alpha(\bw);\bw )}+
\int_0^{+\infty} dx \, x   \frac{f_{X_p,L}(x,L^\alpha(\bw);\bw )}{f(L^\alpha(\bw);\bw )} \\
&
=
\int_{-\infty}^{+\infty} dx \, x f_{X_p | L =L^\alpha(\bw) }(x;\bw )
%=- \EE [ (X_p)_-  - (X_p)_+ | \bw\cdot \bX =L^\alpha]
=
\EE \big[ X_p |L =L^\alpha(\bw)]  ,
\end{align*}
which completes the proof of the second item.
\qed

\end{document}